\documentclass[]{article}
\usepackage{amsmath}
\usepackage{amssymb}
\usepackage{graphicx}

\begin{document}

\title{Smooth quantum gravity: Exotic smoothness and Quantum gravity}

\author{Torsten Asselmeyer-Maluga \\ German Aerospace Center, \\ Rosa-Luxemburg-Str. 2 \\ D-10178 Berlin, Germany \\ {torsten.asselmeyer-maluga@dlr.de}}

\maketitle
\emph{On the occasion of the 80-th birthday of Carl H. Brans}
\begin{abstract}
{Over the last two decades, many unexpected relations between exotic
smoothness, e.g. exotic $\mathbb{R}^{4}$, and quantum field theory
were found. Some of these relations are rooted in a relation to superstring
theory and quantum gravity. Therefore one would expect that exotic
smoothness is directly related to the quantization of general relativity.
In this article we will support this conjecture and develop a new
approach to quantum gravity called \emph{smooth quantum gravity} by
using smooth 4-manifolds with an exotic smoothness structure. In particular
we discuss the appearance of a wildly embedded 3-manifold which
we identify with a quantum state. Furthermore, we analyze this
quantum state by using foliation theory and relate it to an element
in an operator algebra. Then we describe a set of geometric,
non-commutative operators, the skein algebra, which can be used to
determine the geometry of a 3-manifold. This operator algebra can
be understood as a deformation quantization of the classical Poisson
algebra of observables given by holonomies. The structure of this
operator algebra induces an action by using the quantized calculus
of Connes. The scaling behavior of this action is analyzed to obtain
the classical theory of General Relativity (GRT) for large scales. 
This approach has some obvious properties:
there are non-linear gravitons, a connection to lattice gauge field
theory and a dimensional reduction from 4D to 2D. Some cosmological
consequences like the appearance of an inflationary phase are also
discussed. At the end we will get the simple picture that the change
from the standard $\mathbb{R}^{4}$ to the exotic $R^{4}$ is a quantization
of geometry.}
\end{abstract}

\newpage
\tableofcontents

\section{Introduction}

On the 25-th of November in 1915, Einstein presented his field equations, the
basic equations of General Relativity, to the Prussian Academy of
Sciences in Berlin. This equation had a tremendous impact on physics,
in particular on cosmology. The essence of the theory was expressed
by Wheeler by the words: \emph{Spacetime tells matter how to move; matter
tells spacetime how to curve}. Einsteins theory remained unchanged
for about 40 years. Then one started to investigate theories fulfilling
Mach's principle leading to a variable gravitational constant. Brans-Dicke
theory was the first realization of an extended Einstein theory with
variable gravitational constant (Jordans proposal was not widely known).
All experiments are, however, in good agreement  with Einstein's theory
and currently there is no demand to change it.

General relativity (GR) has changed our understanding of space-time.
In parallel, the appearance of quantum field theory (QFT) has modified
our view of particles, fields and the measurement process. The usual
approach for the unification of QFT and GR to a quantum gravity,
starts with a proposal to quantize GR and its underlying structure,
space-time. There is a unique opinion in the community about the relation
between geometry and quantum theory: The geometry as used in GR is
classical and should emerge from a quantum gravity in the limit (Planck's
constant tends to zero). Most theories went a step further and try
to get a space-time from quantum theory. Then, the model of a smooth
manifold is not suitable to describe quantum gravity, but there is
no sign for a discrete space-time structure or higher dimensions in
current experiments \cite{FermiCollab2009}. 
Therefore, we conjecture that the model of spacetime
as a smooth 4-manifold can be used also in a quantum gravity regime, 
but then one has the problem to represent QFT by geometric methods
(submanifolds for particles or fields etc.) as well to quantize GR.
In particular, one must give meaning to the quantum state by geometric
methods. Then one is able to construct the quantum theory without
quantization. Here we implicitly assumed that the quantum state is
real, i.e. the quantum state or the wave function has a real counterpart
and is not a collection of future possibilities representing some
observables. Experiments \cite{Leifer2014,Branciard2014,RDBCWF2015}  
supported this view. Then the wave
function is not merely representing our limited knowledge of a system
but it is in direct correspondence to reality! Then one has to go
the reverse way: one has to show that the quantum state is produced
by the quantization of a classical state. It is, however, not enough to
have a geometric approach to quantum gravity (or the quantum field
theory in general). What are the quantum fluctuations? What is the
measurement process? What is decoherence and entanglement? In principle,
all these questions have to be addressed too. 

Here, the exotic smoothness structure of 4-manifolds can help finding
a way. A lot of work was done in the last decades to fulfill this
goal. It starts with the work of Brans and Randall \cite{BraRan:1993}
and of Brans alone \cite{Brans:99,Brans:94b,Brans:94a} where the special
situation in exotic 4-manifolds (in particular the exotic $\mathbb{R}^{4}$)
was explained. One main result of this time was the {\em Brans conjecture}:
exotic smoothness can serve as an additional source of gravity. I
will not present the whole history where I refer to Carl's article.
Here I will list only some key results which will be used in the following
\begin{itemize}
\item Exotic smoothness is an extra source of gravity (Brans conjecture
is true), see Asselmeyer \cite{Asselmeyer96} for compact manifolds and
S{\l}adkowski \cite{Sla:1999,Sladkowski:2001} for the exotic $\mathbb{R}^{4}$.
Therefore an exotic $\mathbb{R}^{4}$ is always curved and cannot
be flat!
\item The exotic $\mathbb{R}^{4}$ cannot be a globally hyperbolic space
(see \cite{Chernov2012} for instance), i.e. represented by $M\times\mathbb{R}$
for some 3-manifold. Instead it admits complicated foliations \cite{TAMKrol2009}.
Using non-commutative geometry, we are able to study these foliations
(the leaf space) and get relations to QFT. For instance, the von Neumann
algebra of a codimension-one foliation of an exotic $\mathbb{R}^{4}$
must contain a factor of type $I\! I\! I_{1}$ used in local algebraic
QFT to describe the vacuum \cite{TAMKrol2010,TAMKrol2011a,TAMKrol2011d}.
\item The end of $\mathbb{R}^{4}$ (the part extending to infinity) is $S^{3}\times\mathbb{R}$.
If $\mathbb{R}^{4}$ is exotic then $S^{3}\times\mathbb{R}$ admits
also an exotic smoothness structure. Clearly, there is always a topologically
embedded 3-sphere but there is no smoothly embedded one. Let us assume
the well-known hyperbolic metric of the spacetime $S^{3}\times\mathbb{R}$
using the trivial foliation into leafs $S^{3}\times\left\{ t\right\} $
for all $t\in\mathbb{R}$. Now we demand that $S^{3}\times\mathbb{R}$
carries an exotic smoothness structure at the same time. Then we will
get only topologically embedded 3-spheres, the leafs $S^{3}\times\left\{ t\right\} $.
These topologically embedded 3-spheres are also known as wild 3-spheres.
In \cite{TAMKrol2011c}, we presented a relation to quantum
D-branes. Finally we proved in \cite{TAMKrol2013} that the
deformation quantization of a tame embedding (the usual embedding)
is a wild embedding%
\footnote{A wild embedding is a topological embedding $I:N\to M$ so that the
image $I(N)\subset M$ is an infinite polyhedron or the triangulation
needs always infinitely many simplices. %
}. Furthermore we obtained a geometric interpretation of quantum states:
wild embedded submanifolds are quantum states. Importantly, this construction
depends essentially on the continuum, because wild embedded submanifolds
admit always infinite triangulations.
\item For a special class of compact 4-manifolds we showed in \cite{AsselmeyerRose:2012}
that exotic smoothness can generate fermions and gauge fields using
the so-called knot surgery of Fintushel and Stern \cite{FinSte:1998}.
In the paper \cite{TAMBrans2015} we presented an approach
using the exotic $\mathbb{R}^{4}$ where the matter can be generated
(like in QFT).
\item The path integral in quantum gravity is dominated by the exotic smoothness
contribution (see \cite{TAM2010,Duston2010,Pfeiffer:2004} or by using
string theory \cite{TAMKrol2010ICM}).
\end{itemize}
The paper is organized as follows. In the following three sections
we will explain exotic 4-manifolds and motivate the whole approach
by using the path integral for the Einstein-Hilbert action. Here we
will also present how to couple the matter and gauge fields to this
theory. For a 4-manifold, there are two main invariants the Euler
and Pontrjagin class which determine the main topological invariant
of a 4-manifold, the intersection form. In section \ref{sec:action-induced-by-top},
we will obtain the Einstein-Hilbert and Holst action by using these
two classes. At the first view, this section is a little bit isolated
from the previous and subsequent sections but we will use this result
later during the study of the scaling. In the main section \ref{sec:Wild-embeddings},
we will construct the foliation of an exotic $\mathbb{R}^{4}$ of
codimension (equivalent to a Lorentz structure). Following Connes,
\cite{Connes1984} the leaf space is an operator algebra constructed from the geometrical
information of the foliation (holonomy groupoid). This operator algebra is a factor
$I\! I\! I$ von Neumann algebra and we will use the Tomita-Takesaki modular theory
to uncover the structure of the foliation. 
It is not the first time that this factor was used for
quantum gravity and we refer to the paper \cite{BertozziniConti2010} 
for a nice application. States in this operator
algebra are represented by equivalence classes of knotted curves (element
of the Kauffman bracket skein module). The reconstruction of the spatial
space from the states gives a wild embedded 3-sphere as geometrical
representation of the state. Surprisingly, it fits with the properties
of the exotic $\mathbb{R}^{4}$. If one introduces a global foliation
of the exotic $\mathbb{R}^{4}$ by a global time then one obtains
a foliation into wild embedded 3-spheres. In contrast, if one uses
a local but complicate foliation then this wild object can be omitted
and one obtains a state given by a finite collection of knotted curves.
Interestingly, the operator algebra can be understood as observable
algebra given by a deformation quantization 
(Turaev-Drinfeld quantization \cite{Turaev1989,Turaev1991})
of the classical observable algebra (Poisson algebra of holonomies
a la Goldman \cite{Goldman1984}). In section \ref{sec:Action-at-the-quantum-level},
we will use the splitting of the operator algebra (\ref{eq:crossed-product-factor-III})
given by Tomita-Takesaki modular theory to introduce the dynamics
(see Connes and Rovelli \cite{ConnesRovelli1994} with similar ideas). Finally we will
obtain a quantum action (\ref{eq:quantum-action}) in the quantized
calculus of Connes \cite{Connes94}. Then the scaling behavior is studied in the next
section. For large scales, the action can be interpreted as a non-linear
sigma model. The renormalization group (RG) flow analysis \cite{Friedan1980} 
gives the Einstein equations for
large scales. The short-scale analysis is much more involved, yielding
for small fluctuations the Einstein-Hilbert action and a non-minimally
coupled scalar field. In particular, we will obtain a $(2+\epsilon)-$dimensional
fractal structure.  In section \ref{sec:Some-Properties} we will
present some direct consequences of this approach: the nonlinear graviton \cite{Penrose1976},
a relation to lattice gauge field theory with a discussion of discreteness
and the appearance of dimensional reduction from 4D to 2D. In section \ref{sec:Where-does-fluctuation} we will discuss the answer
to a fundamental question: where does the quantum fluctuations come
from? The main result of this section can be written as: \emph{The
set of canonical pairs (as measurable variables in the theory) forms
a fractal subset of the space of all holonomies. Then we can only
determine the initial condition up to discrete value (given by the
canonical pair) and the chaotic behavior of the foliation (i.e. the
Anosov flow) makes the limit not predictable.} This interesting result
is followed by a section where we will discuss the collapse of the
wavefunction by the gravitational interaction by calculating the minimal
decoherence time. Furthermore we will discuss entanglement and the
measurement process. In section \ref{sec:Some-Implications-for.cosmology}
we will list our work in cosmology which uses partly the results of
this paper. In the last section \ref{sec:Conclusion-and-Open-questions},
we will discuss some consequences and open questions. 
Some mathematical prerequisites are presented in three 
appendices.

This article is dedicated to my only teacher, Carl H. Brans for 20
years of collaboration and friendship. He is the founder of this research
area. We had and will have many interesting discussions. Carl always
asked the right question and put the finger on many open points. During
the 7 years of writing our book, we had a very fruitful collaboration
and I learned so much to complete even this work. Carl, I hope for
many discussions with you in the future. I'm very glad to count on your
advice. Happy Birthday!

\section{What is Exotic Smoothness?}
Why am I going to concentrate on a concept like exotic smoothness? Einstein
used the equivalence principle as a key principle in the development
of general relativity. Every gravitational field can be locally eliminated
by acceleration. Then, the spacetime is locally modeled as subsets
of the flat $\mathbb{R}^{4}$ or the equivalence principle enforces
us to use the concept of a manifold for spacetime. Together with the
smoothness of the dynamics (usage of differential equations), we obtain
a smooth 4-manifold as model for the spacetime in agreement with the
current experimental situation. A manifold consists of charts and
transition functions forming an atlas which covers the manifold completely.
\emph{The smooth atlas is called the smoothness structure} of the
manifold. It was an open problem for a long time whether every topological
manifold admits a unique smooth atlas. In 1957, Milnor found the first
counterexample: the construction of a 7-sphere with at least 8 different
smoothness structures. Later it was shown that all manifolds of dimension
larger than 4 admit only a finite number of distinct smoothness structures.
The real breakthrough for 4-manifolds came in the 80's where one constructed infinitely many different smoothness structures for many compact 4-manifolds
(countably infinite) and for many non-compact 4-manifolds (uncountably
infinite) including the $\mathbb{R}^{4}$. In all dimensions smaller
than four, there is only one smoothness structure (up to diffeomorphisms),
the standard structure. The standard $\mathbb{R}^{4}$ is simply characterized
by the unique property to split smoothly like $\mathbb{R}^{4}=\mathbb{R}\times\mathbb{R}\times\mathbb{R}\times\mathbb{R}$.
All other distinct smoothness structures are called \emph{exotic smoothness}
structures. These structures are different, nonequivalent, smooth
descriptions of the same topological manifold, a different atlas of
charts. In case of the exotic $\mathbb{R}^{4}$, the difference is
tremendous: the standard $\mathbb{R}^{4}$ needs one chart (and every
other description can be reduced to it) whereas every known exotic
$\mathbb{R}^{4}$ admits infinitely many charts (which cannot be reduced
to a simpler description). So, the spacetime exhibits a much larger
complexity by using an exotic smoothness structure, but why is dimension
4 so special? There is a good description in \cite{Fre:1984} and I
will give a short account now. At first we have to discuss the question:
how do I build an atlas for a smooth manifold? The answer is given
by considering the construction of diffeomorphisms. Every diffeomorphism
is locally given by the solution of $\dot{\vec{x}}=-\nabla f(\vec{x})$
for a real function $f$ over the manifold. The fixed points of this
equation are the critical points of $f$. In case of isolated critical
points, one can reproduce the structure of the manifold (this is called
Morse theory). Every critical point leads to the attachment of a handle,
a submanifold like $D^{n-k}\times D^{k}$, i.e. the $k-$handle (where
$D^{k}$ is the $k-$disk). In many cases, the corresponding structure
of the manifold, the handle body, can be very complicated but there
are rules (handle sliding) to simplify them. In all dimensions except
dimension 4. Therefore, two handle bodies can be described by the same 4-manifold topologically but differ in the smooth description. 

\section{The Main Example: Exotic $\mathbb{R}^{4}$}

One of the most surprising aspects of exotic smoothness is the existence
of exotic $\mathbb{R}^{4}$'s. In all other dimensions \cite{Sta:1962},
the Euclidean space $\mathbb{R}^{n}$ with $n\not=4$ admits a unique
smoothness structure, up to diffeomorphisms. Beginning with the first
examples \cite{Gom:1983}, Taubes \cite{Tau:87} and Freedman/DeMichelis
\cite{DeMichFreedman1992} constructed countably many large and small
exotic $\mathbb{R}^{4}$'s, respectively. A small exotic $\mathbb{R}^{4}$
embeds smoothly in the 4-sphere whereas a large exotic $\mathbb{R}^{4}$
cannot be embedded in that way. For the following we need some simple definitions:
the connected sum $\#$ and the boundary connected sum $\natural$
of manifolds. Let $M,N$ be two $n$-manifolds with boundaries $\partial M,\partial N$.
The \emph{connected sum} $M\#N$ is the procedure of cutting out a
disk $D^{n}$ from the interior $int(M)\setminus D^{n}$ and $int(N)\setminus D^{n}$
with the boundaries $S^{n-1}\sqcup\partial M$ and $S^{n-1}\sqcup\partial N$,
respectively, and gluing them together along the common boundary component
$S^{n-1}$. The boundary $\partial(M\#N)=\partial M\sqcup\partial N$
is the disjoint sum of the boundaries $\partial M,\partial N$. The
\emph{boundary connected sum} $M\natural N$ is the procedure of cutting
out a disk $D^{n-1}$ from the boundary $\partial M\setminus D^{n-1}$
and $\partial N\setminus D^{n-1}$ and gluing them together along
$S^{n-2}$ of the boundary. Then the boundary of this sum $M\natural N$
is the connected sum $\partial(M\natural N)=\partial M\#\partial N$
of the boundaries $\partial M,\partial N$.

\subsection{Large Exotic $\mathbb{R}^{4}$}

Large exotic $\mathbb{R}^{4}$ can be constructed using the failure
to arbitrarily split a compact, simply-connected 4-manifold. For every
topological 4-manifold one knows how to split this manifold {\em
topologically} into simpler pieces using the work of Freedman \cite{Fre:82}.
Donaldson \cite{Don:1983}, however, that some of these 4-manifolds
do not exist as smooth 4-manifolds. This contradiction between the
continuous and the smooth case produces the first examples of exotic
$\mathbb{R}^{4}$. Below we discuss one of these examples.

One starts with a compact, simply-connected 4-manifold $X$ classified
by the intersection form \cite{Fre:82}
\[
Q_{X}:H_{2}(X,\mathbb{Z})\times H_{2}(X,\mathbb{Z})\to\mathbb{Z}
\]
 a quadratic form over the second integer homology group. In the first
construction of a large exotic $\mathbb{R}^{4}$, one starts with
the K3 surface as 4-manifold having the intersection form
\begin{equation}
Q_{K3}=E_{8}\oplus E_{8}\oplus(\oplus_{3}\left(\begin{array}{cc}
0 & 1\\
1 & 0
\end{array}\right)):=2E_{8}\oplus3H\label{eq:intersection-K3}
\end{equation}
 with the the matrix $E_{8}$: 
\[
E_{8}=\left[\begin{array}{cccccccc}
2 & 1 & 0 & 0 & 0 & 0 & 0 & 0\\
1 & 2 & 1 & 0 & 0 & 0 & 0 & 0\\
0 & 1 & 2 & 1 & 0 & 0 & 0 & 0\\
0 & 0 & 1 & 2 & 1 & 0 & 0 & 0\\
0 & 0 & 0 & 1 & 2 & 1 & 0 & 1\\
0 & 0 & 0 & 0 & 1 & 2 & 1 & 0\\
0 & 0 & 0 & 0 & 0 & 1 & 2 & 0\\
0 & 0 & 0 & 0 & 1 & 0 & 0 & 2
\end{array}\right].
\]
 The work of Donaldson \cite{Don:1983} shows that a closed, \textbf{smooth},
simply-connected, compact 4-manifold $X_{E_{8}\oplus E_{8}}$ with
intersection form $E_{8}\oplus E_{8}$ does not exist. Freedman
\cite{Fre:82} showed, however, that there is a topological splitting
\begin{equation}
K3=X_{E_{8}\oplus E_{8}}\#\left(\#_{3}(S^{2}\times S^{2})\right)\label{eq:splitting-K3}
\end{equation}
 with the $m-$times connected sum $\#_{m}$ (see above) which fails
to be smooth. This splitting means that we glue together the two manifolds
$\#_{3}(S^{2}\times S^{2})\setminus D^{4}$ and $X_{E_{8}\oplus E_{8}}\setminus D^{4}$
along the common boundary $S^{3}=\partial D^{4}$ ($D^{4}$ is the
4-disk or 4-ball). Now we define the interior $X=\#_{3}(S^{2}\times S^{2})\setminus Int\, D^{4}$.
The splitting (\ref{eq:splitting-K3}) gives a way to represent the
$3H$ part of the intersection form (\ref{eq:intersection-K3}) by
using $X$ but that fails smoothly. So, choosing a topological splitting
\begin{eqnarray*}
K3 & = & X_{E_{8}\oplus E_{8}}\#\left(\#_{3}(S^{2}\times S^{2})\right)\\
 & = & \left(X_{E_{8}\oplus E_{8}}\setminus D^{4}\right)\cup\left(S^{3}\times[0,1]\right)\cup\left(\#_{3}(S^{2}\times S^{2})\setminus D^{4}\right)
\end{eqnarray*}
 gives a $S^{3}\times[0,1]$ inside the $K3$. The interior of $S^{3}\times[0,1]$
defines a manifold $S^{3}\times[0,1)$ glued to a (topological) 4-disk
$D^{4}\subset\#_{3}(S^{2}\times S^{2})\setminus D^{4}$ along the
common boundary, i.e. $W=D^{4}\cup S^{3}\times[0,1)$ topologically.
$W$ is homeomorphic to $\mathbb{R}^{4}$ but the non-existence of
the smooth splitting implies that it is an exotic $\mathbb{R}^{4}$
and there is no smooth embedded $S^{3}$ (otherwise the topological
splitting is smooth). This failure for a smooth embedding implies
also that such exotic $\mathbb{R}^{4}$'s do not embed in the 4-sphere,
i.e. it is a large exotic $\mathbb{R}^{4}$. The details of the construction
can be found in our book \cite{AsselmeyerBrans2007} (section 8.4).

Gompf \cite{Gom:85} introduced an important tool for finding new
exotic ${{\mathbb{R}}^{4}}$ from others, the end-sum $\natural_{e}$.
Let $R,R'$ be two topological ${{\mathbb{R}}^{4}}$'s. The end-sum
$R\natural_{e}R'$ is defined as follows: Let $\gamma:[0,\infty)\to R$
and $\gamma':[0,\infty)\to R'$ be smooth properly embedded rays with
tubular neighborhoods $\nu\subset R$ and $\nu'\subset R'$, respectively.
For convenience, identify the two semi-infinite intervals with $[0,1/2),$
and $(1/2,1]$ leading to diffeomorphisms, $\phi:\nu\to[0,1/2)\times\mathbb{R}^{3}$
and $\phi':\nu'\to(1/2,1]\times\mathbb{R}^{3}$. Then define 
\[
R\natural_{e}R'=R\cup_{\phi}I\times{\mathbb{R}}^{3}\cup_{\phi'}R'
\]
 as the end sum of $R$ and $R'$. With a little checking, it is easy
to see that this construction leads to $R\natural_{e}R'$ as another
topological $\mathbb{R}^{4}.$ However, if $R,R'$ are themselves
exotic, then so will $R\natural_{e}R'$ and in fact, it will be a
``new'' exotic manifold, since it will not be diffeomorphic to either
$R$ or $R'$. Gompf used this technique to construct a class of exotic
${{\mathbb{R}}^{4}}$'s none of which can be embedded smoothly in
the standard ${{\mathbb{R}}^{4}}$.

By an extension of Donaldson theory for a special class of open 4-manifolds,
so-called end-periodic 4-manifolds, Taubes \cite{Tau:87} gives
a continuous family of exotic $\mathbb{R}^{4}$ which was extended
by Gompf to a continuous 2-parameter family $R_{s,t}$.

\subsection{Small Exotic $\mathbb{R}^{4}$\label{sub:Small-exotic}}

Small exotic $\mathbb{R}^{4}$'s are again the result of anomalous
smoothness in 4-dimensional topology but of a different kind than
for large exotic $\mathbb{R}^{4}$'s. In 4-manifold topology \cite{Fre:82},
a homotopy-equivalence between two compact, closed, simply-connected
4-manifolds implies a homeomorphism between them (a so-called h cobordism), but Donaldson \cite{Don:87} provided the first smooth counterexample,
i.e. both manifolds are generally not diffeomorphic to each other.
The failure can be localized in some contractible submanifold (Akbulut
cork) so that an open neighborhood of this submanifold is a small
exotic $\mathbb{R}^{4}$. The whole procedure implies that this exotic
$\mathbb{R}^{4}$ can be embedded in the 4-sphere $S^{4}$. Below
we discuss the details for one of these examples.

In 1975 Casson (Lecture 3 in \cite{Cas:73}) described a smooth 5-dimensional
h-cobordism between compact 4-manifolds and showed that they ``differ''
by two proper homotopy ${{\mathbb{R}}^{4}}$'s (see below). Freedman
knew, as an application of his proper h-cobordism theorem, that the
proper homotopy ${{\mathbb{R}}^{4}}$'s were ${{\mathbb{R}}^{4}}$.
After hearing about Donaldson's work in March 1983, Freedman realized
that there should be exotic ${{\mathbb{R}}^{4}}$'s and, to find one,
he produced the second part of the construction below involving the
smooth embedding of the proper homotopy ${{\mathbb{R}}^{4}}$'s in
$S^{4}$. Unfortunately, it was necessary to have a compact counterexample
to the smooth h-cobordism conjecture, and Donaldson did not provide
this until 1985 \cite{Don:87}. The idea of the construction is simply
given by the fact that every such smooth h-cobordism between non-diffeomorphic
4-manifolds can be written as a product cobordism except for a compact
contractible sub-h-cobordism $V$, the Akbulut cork. An open subset
$U\subset V$ homeomorphic to $[0,1]\times{{\mathbb{R}}^{4}}$ is
the corresponding sub-h-cobordism between two exotic ${{\mathbb{R}}^{4}}$'s.
These exotic ${{\mathbb{R}}^{4}}$'s are called ribbon ${{\mathbb{R}}^{4}}$'s.
They have the important property of being diffeomorphic to open subsets
of the standard ${{\mathbb{R}}^{4}}$. That stands in contrast to
the previous defined examples of Kirby, Gompf and Taubes.

To be more precise, consider a pair $(X_{+},X_{-})$ of homeomorphic,
smooth, closed, simply-connected 4-manifolds. The transformation from
$X_{-}$ to $X_{+}$ visualized by a h-cobordism can be described
by the following construction. \\
 \emph{Let $W$ be a smooth h-cobordism between closed, simply connected
4-manifolds $X_{-}$ and $X_{+}$. Then there is an open subset $U\subset W$
homeomorphic to $[0,1]\times{{\mathbb{R}}^{4}}$ with a compact subset
$K\subset U$ such that the pair $(W\setminus K,U\setminus K)$ is
diffeomorphic to a product $[0,1]\times(X_{-}\setminus K,U\cap X_{-}\setminus K)$.
The subsets $R_{\pm}=U\cap X_{\pm}$ (homeomorphic to ${{\mathbb{R}}^{4}}$)
are diffeomorphic to open subsets of ${{\mathbb{R}}^{4}}$. If $X_{-}$
and $X_{+}$ are not diffeomorphic, then there is no smooth 4-ball
in $R_{\pm}$ containing the compact set $Y_{\pm}=K\cap R_{\pm}$,
so both $R_{\pm}$ are exotic ${{\mathbb{R}}^{4}}$'s.} \\
 Thus, remove a certain contractible, smooth, compact 4-manifold $Y_{-}\subset X_{-}$
(called an Akbulut cork) from $X_{-}$, and re-glue it by an involution
of $\partial Y_{-}$, i.e. a diffeomorphism $\tau:\partial Y_{-}\to\partial Y_{-}$
with $\tau\circ\tau=Id$ and $\tau(p)\not=\pm p$ for all $p\in\partial Y_{-}$.
This argument was modified above so that it works for a contractible
{\em open} subset $R_{-}\subset X_{-}$ with similar properties,
such that $R_{-}$ will be an exotic ${{\mathbb{R}}^{4}}$ if $X_{+}$
is not diffeomorphic to $X_{-}$. Furthermore $R_{-}$ lies in a compact
set, i.e. a 4-sphere or $R_{-}$ is a small exotic $\mathbb{R}^{4}$.
In the next subsection we will see how this results in the construction
of handle bodies of exotic ${{\mathbb{R}}^{4}}$. In \cite{DeMichFreedman1992}
Freedman and DeMichelis constructed also a continuous family of small
exotic $\mathbb{R}^{4}$.

\subsection{Main Property of (Small) Exotic $\mathbb{R}^{4}$}

One of the characterizing properties of an exotic $\mathbb{R}^{4}$
(all known examples) is the existence of a compact subset $K\subset R^{4}$
which cannot be surrounded by any smoothly embedded 3-sphere (and
homology 3-sphere bounding a contractible, smooth 4-manifold). Let
$\mathbf{R}^{4}$ be the standard $\mathbb{R}^{4}$ (i.e. $\mathbf{R}^{4}=\mathbb{R}^{3}\times\mathbb{R}$
smoothly) and let $R^{4}$ be a small exotic $\mathbb{R}^{4}$ with
compact subset $K\subset R^{4}$ which cannot be surrounded by a smoothly
embedded 3-sphere. Then every completion $\overline{N(K)}$ of an
open neighborhood $N(K)\subset R^{4}$ is not bounded by a 3-sphere
$S^{3}\not=\partial\overline{N(K)}$. However, $R^{4}$ is a small exotic
$\mathbb{R}^{4}$ and there is a smooth embedding $E:R^{4}\to\mathbf{R}^{4}$
in the standard $\mathbb{R}^{4}$. Then the completion of the image
$\overline{E(R^{4})}$ has the boundary $S^{3}=\partial\overline{E(R^{4})}$
as subset of $\mathbf{R}^{4}$. So, we have the strange situation
that an open subset of the standard $\mathbf{R}^{4}$ represents a
small exotic $R^{4}$. In case of the large exotic $\mathbb{R}^{4}$,
the situation is much more complicated. A large exotic $\mathbb{R}^{4}$
does not embed in any smooth 4-manifold which is simpler than the manifold
used for the construction of this exotic $\mathbb{R}^{4}$. Above
we considered the example of a large exotic $\mathbb{R}^{4}$ constructed
from a K3 surface. Therefore this large exotic $\mathbb{R}^{4}$ embeds
in the K3 surface but not in simpler 4-manifolds like $\mathbb{C}P^{2}$.

\subsection{Handle decomposition of the small exotic $\mathbb{R}^{4}$ and Casson
handles\label{sub:Handle-decomposition-Casson-handle}}

As of now, we only know of exotic $\mathbb{R}^{4}$'s represented
by an infinite number of coordinate patches. This naturally makes
it difficult to provide an explicit description of a metric. However,
in \cite{TAMBrans2012}, a suggestion to overcome this limitation
is provided by the consideration of periodic explicitly described
coordinate patches making use of more complex pieces, so-called handles,
and even more complex gluing maps. Then one also gets infinite structures
of handles but with a clear picture: the coordinate patches have a
periodic structure.

\textbf{Handles} Every 4-manifold can be decomposed using standard
pieces such as $D^{k}\times D^{4-k}$, the so-called $k$-handle attached
along $\partial D^{k}\times D^{4-k}$ to the $0-$handle $D^{0}\times D^{4}=D^{4}$.
In the following we need two possible cases: the 1-handle $D^{1}\times D^{3}$
and the 2-handle $D^{2}\times D^{2}$. These handles are attached
along their boundary components $S^{0}\times D^{3}$ or $S^{1}\times D^{2}$
to the boundary $S^{3}$ of the $0-$handle $D^{4}$ (see \cite{GomSti:1999}
for the details). The attachment of a 2-handle is defined by a map
$S^{1}\times D^{2}\to S^{3}$, the embedding of a circle $S^{1}$
into the 3-sphere $S^{3}$, i.e. a knot. This knot into $S^{3}$ can
be thickened (to get a knotted solid torus). The important fact for
our purposes is the freedom to twist this knotted solid torus (so-called
Dehn twist). The (integer) number of these twists (with respect to
the orientation) is called the framing number or the framing. Thus
the gluing of the 2-handle on $D^{4}$ can be represented by a knot
or link together with an integer framing. The simplest example is
the unknot with framing $\pm1$ representing the complex projective
space $\mathbb{C}P^{2}$ or with reversed orientation $\overline{\mathbb{C}P}^{2}$,
respectively. The 1-handle will be glued by the map of $S^{0}\times D^{3}\to S^{3}$
represented by two disjoint solid 2-spheres $D^{3}$. Akbulut \cite{AkbKir:79}
introduced another description. He observed that a 1-handle is something
like a cut-out 2-handle with a fixed framing. We remark that all details
can be found in \cite{GomSti:1999}. Now we are ready to discuss the
handle body decomposition of an exotic $\mathbb{R}^{4}$ by Bizaca
and Gompf \cite{BizGom1996}.

\textbf{Handle decomposition of small exotic $\mathbb{R}^{4}$} First
it is very important to notice that the exotic $\mathbb{R}^{4}$ is
the \textbf{interior} of the handle body described below (since the
handle body has a non-null boundary and is compact). The construction
of the handle body can be divided into two parts. The first part is
a submanifold consisting of a pair of a 1- and a 2-handle. This pair
can be canceled topologically by using a Casson handle and we obtain
the topological 4-disk $D^{4}$ with $\mathbb{R}^{4}$ as interior.
This submanifold is a smooth 4-manifold with a boundary that can be
covered by a finite number of charts. The smoothness structure
of the exotic $\mathbb{R}^{4}$, however, depends mainly on the infinite Casson handle.

\textbf{Casson handle} Now consider the Casson handle and its construction
in more detail. Briefly, a Casson handle $CH$ is the result of attempts
to embed a disk $D^{2}$ into a 4-manifold. In most cases this attempt
fails and Casson \cite{Cas:73} looked for a substitute, which is
now called a Casson handle. Freedman \cite{Fre:82} showed that every
Casson handle $CH$ is homeomorphic to the open 2-handle $D^{2}\times\mathbb{R}^{2}$
but in nearly all cases it is not diffeomorphic to the standard handle
\cite{Gom:84,Gom:89}. The Casson handle is built by iteration, starting
from an immersed disk in some 4-manifold $M$, i.e. an injective smooth
map $D^{2}\to M.$ Every immersion $D^{2}\to M$ is an embedding except
on a countable set of points, the double points. One can kill one
double point by immersing another disk into that point. These disks
form the first stage of the Casson handle. By iteration one can produce
the other stages. Finally consider not the immersed disk but rather
a tubular neighborhood $D^{2}\times D^{2}$ of the immersed disk including
each stage. The union of all neighborhoods of all stages is the Casson
handle $CH$. So, there are two input data involved with the construction
of a $CH$: the number of double points in each stage and their orientation
$\pm$. Thus we can visualize the Casson handle $CH$ by a tree: the
root is the immersion $D^{2}\to M$ with $k$ double points, the first
stage forms the next level of the tree with $k$ vertices connected
with the root by edges etc. The edges are evaluated using the orientation
$\pm$. Every Casson handle can be represented by such an infinite
tree. The Casson handle $CH(R_{+})$ having an immersed disk with
one (positively oriented) self-intersection (or double point) is the
simplest Casson handle represented by the simplest tree $T_{+}$ having
one vertex in each level connected by one edge with evaluation $+$.

\subsection{Small Exotic $\mathbb{R}^{4}$ as a Sequence of 3-Manifolds\label{sub:Small-exotic-R4-sequnce-3MF}}

One of the characterizing properties of an exotic $\mathbb{R}^{4}$
(all known examples) is the existence of a compact subset $K\subset R^{4}$
which cannot be surrounded by any smoothly embedded 3-sphere (and
homology 3-sphere bounding a contractible, smooth 4-manifold). Let
$\mathbf{R}^{4}$ be the standard $\mathbb{R}^{4}$ (i.e. $\mathbf{R}^{4}=\mathbb{R}^{3}\times\mathbb{R}$
smoothly) and let $R^{4}$ be a small exotic $\mathbb{R}^{4}$ with
compact subset $K\subset R^{4}$ which cannot be surrounded by a smoothly
embedded 3-sphere. Then every completion $\overline{N(K)}$ of an
open neighborhood $N(K)\subset R^{4}$ is not bounded by a 3-sphere
$S^{3}\not=\partial\overline{N(K)}$, but $R^{4}$ is a small exotic
$\mathbb{R}^{4}$ and there is a smooth embedding $E:R^{4}\to\mathbf{R}^{4}$
in the standard $\mathbb{R}^{4}$. Then the completion of the image
$\overline{E(R^{4})}$ has the boundary $S^{3}=\partial\overline{E(R^{4})}$
as subset of $\mathbf{R}^{4}$. So, we have the strange situation
that an open subset of the standard $\mathbf{R}^{4}$ represents a
small exotic $R^{4}$.

\begin{figure}
\centering \resizebox{0.25\textwidth}{!}{ \includegraphics{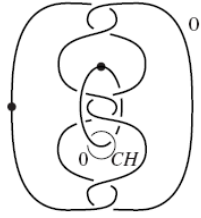}
} 
\caption{Link picture for the compact subset $K$ \label{fig:link-picture-for-K}}
\end{figure}

Now we will describe $R^{4}$. Historically it was constructed by
using a counterexample of the smooth h-cobordism theorem \cite{Don:87,BizGom1996}.
Then the compact subset $K$ is given by a non-canceling 1-/2-handle
pair. The attachment of a Casson handle $CH$ cancels this pair topologically.
Then one obtains the 4-disk $D^{4}$ with interior $\mathbf{R}^{4}$, 
but this cancellation of the 1/2-handle pair cannot be done smoothly
and one obtains a small exotic $R^{4}$ which is schematically given
by $R^{4}=K\cup CH$. Remember $R^{4}$ is a small exotic $\mathbb{R}^{4}$,
i.e. $R^{4}$ is embedded into the standard $\mathbf{R}^{4}$ by definition.
The completion $\bar{R}^{4}$ of $R^{4}\subset\mathbf{R}^{4}$ has
a boundary given by the 3-manifold $Y_{r}$. There is also the possibility
to construct $Y_{r}$ directly as the limit $n\to\infty$ of a sequence
$\left\{ Y_{n}\right\} $ of 3-manifolds. To construct this sequence
of 3-manifolds \cite{Ganzel2006}, one can use the Kirby calculus,
i.e. one represents the compact subset $K$ by 1- and 2-handles pictured
by a link say $L_{K}$ where the 1-handles are represented by a dot
(so that surgery along this link gives $K$) \cite{GomSti:1999}.
Then one attaches a Casson handle to this link \cite{BizGom1996}.
As an example see Figure \ref{fig:link-picture-for-K}. 

\begin{figure}
\begin{centering}
\resizebox{0.7\textwidth}{!}{ \includegraphics{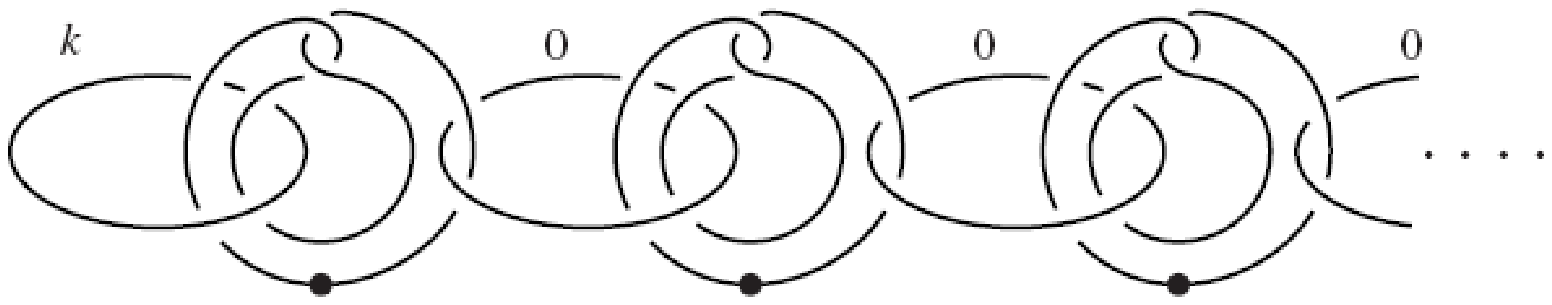}\qquad{}\qquad{}\qquad{}\includegraphics{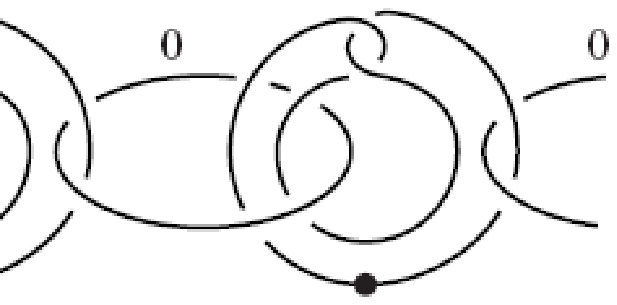}
} 
\par\end{centering}
\caption{Building block of every Casson handle (right) and the simplest Casson
handle (left).\label{fig:building-block-simplest-CH}}
\end{figure}

The Casson handle is given by a sequence of Whitehead links (where
the unknotted component has a dot) which are linked according to the
tree (see the right figure of Figure \ref{fig:building-block-simplest-CH}
for the building block and the left figure for the simplest Casson
handle given by the unbranched tree). 

\begin{figure}
\centering \resizebox{0.25\textwidth}{!}{ \includegraphics{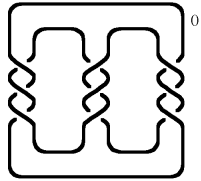}
}
\caption{Pretzel knot $(-3,3,-3)$ or the knot $9_{46}$ in Rolfson notation
producing the 3-manifold $Y_{1}$ by $0-$framed Dehn surgery.\label{fig:pretzel-knot}}
\end{figure}

For the construction of a 3-manifold which surrounds the compact $K$,
one considers $n-$stages of the Casson handle and transforms the
diagram to a real link (the dotted components are changed to usual
components with framing $0$). By handle manipulations one obtains
a knot so that the $n$-th (untwisted) Whitehead double of this knot
represents the desired 3-manifold (by using surgery). Then our example
in Figure \ref{fig:link-picture-for-K} will result in the $n$-th
untwisted Whitehead double of the pretzel knot $(-3,3,-3)$, Figure
\ref{fig:pretzel-knot} (see \cite{Ganzel2006} for the handle manipulations).

 Then this sequence of 3-manifolds 
\[
Y_{1}\to Y_{2}\to\cdots\to Y_{\infty}=Y_{r}
\]
characterizes the exotic smoothness structure of $R^{4}$. The limit
of this sequence $n\to\infty$ gives a wild embedded 3-manifold $Y_{r}$
whose physical relevance will be explained later.

\section{Motivation: Path Integral Contribution by Exotic Smoothness} \label{sec:pathint}

Here, we will motivate the appearance of exotic smoothness by discussing
the path integral for the Einstein-Hilbert action. For simplicity,
we consider general relativity without matter (using the notation
of topological QFT). Space-time is a smooth oriented 4-manifold $M$
which is non-compact and without boundary. From the formal point of
view (no divergences of the metric) one is able to define a boundary
$\partial M$ at infinity. The classical theory is the study of the
existence and uniqueness of (smooth) metric tensors $g$ on $M$ satisfying
the Einstein equations subject to suitable boundary conditions. In
the first order Hilbert\textendash{}Palatini formulation, one specifies
an $SO(1,3)$-connection $A$ together with a cotetrad field $e$
rather than a metric tensor. Fixing $A|_{\partial M}$ at the boundary,
one can derive first order field equations in the interior (now called
\emph{bulk}) which are equivalent to the Einstein equations provided
that the cotetrad is non-degenerate. The theory is invariant under
space-time diffeomorphisms $M\to M$. In the particular case of the
space-time $M=S^{3}\times\mathbb{R}$ (topologically), we have to
consider smooth 4-manifolds $M_{i,f}$ as parts of $M$ whose boundary
$\partial M_{i,f}=\Sigma_{i}\sqcup\Sigma_{f}$ is the disjoint union
of two smooth 3-manifolds $\Sigma_{i}$ and $\Sigma_{f}$ to which
we associate Hilbert spaces $\mathcal{H}_{j}$ of 3-geometries, $j=i,f$.
These contain suitable wave functionals of connections $A|_{\Sigma_{j}}$
. We denote the connection eigenstates by $|A|_{\Sigma_{j}}\rangle$.
The path integral,
\begin{equation}
\langle A|_{\Sigma_{f}}|T_{M}|A|_{\Sigma_{i}}\rangle=\intop_{A|\partial M_{i,f}}DA\, De\,\exp\left(\frac{i}{\hbar}S_{EH}[e,A,M_{i,f}]\right)\label{eq:path-integral}
\end{equation}
is the sum over all connections $A$ matching $A|_{\partial M_{i,f}}$,
and over all $e$. It yields the matrix elements of a linear map $T_{M}:\mathcal{H}_{i}\to\mathcal{H}_{f}$
between states of 3-geometry. Our basic gravitational variables will
be cotetrad $e_{a}^{I}$ and connection $A_{a}^{IJ}$ on space-time
$M$ with the index $a$ to present it as 1-forms and the indices
$I,J$ for an internal vector space $V$ (used for the representation
of the symmetry group). Cotetrads $e$ are \textquoteleft{}square-roots\textquoteright{}
of metrics and the transition from metrics to tetrads is motivated
by the fact that tetrads are essential if one is to introduce spinorial
matter. $e_{a}^{I}$ is an isomorphism between the tangent space $T_{p}(M)$
at any point $p$ and a fixed internal vector space $V$ equipped
with a metric $\eta_{IJ}$ so that $g_{ab}=e_{a}^{I}e_{b}^{J}\eta_{IJ}$.
Here we used the action
\begin{equation}
S_{EH}[e,A,M_{i,f},\partial M_{i,f}]=\intop_{M_{i,f}}\epsilon_{IJKL}(e^{I}\wedge e^{J}\wedge\left(dA+A\wedge A\right)^{KL})+\intop_{\partial M_{i,f}}\epsilon_{IJKL}(e^{I}\wedge e^{J}\wedge A^{KL})\label{eq:action-with-boundary}
\end{equation}
in the notation of \cite{Ashtekar08,Ashtekar08a}. The boundary term
$\epsilon_{IJKL}(e^{I}\wedge e^{J}\wedge A^{KL})$ is equal to twice
the trace over the extrinsic curvature (or the mean curvature). For
fixed boundary data, (\ref{eq:path-integral}) is a diffeomorphism
invariant in the bulk. If $\Sigma_{i}=\Sigma_{f}$ are diffeomorphic,
we can identify $\Sigma=\Sigma_{i}=\Sigma_{f}$ and $\mathcal{H}=\mathcal{H}_{i}=\mathcal{H}_{f}$
i.e. we close the manifold $M_{i,f}$ by identifying the two boundaries
to get the closed 4-manifold $M'$. Provided that the trace over $\mathcal{H}$
can be defined, the partition function, 
\begin{equation}
Z(M')=tr_{\mathcal{H}}T_{M}=\int DA\, De\,\exp\left(\frac{i}{\hbar}S_{EH}[e,A,M,\partial M]\right)\label{eq:path-integral-1}
\end{equation}
where the integral is now unrestricted, is a dimensionless number
which depends only on the diffeomorphism class of the smooth manifold
$M'$. In case of the manifold $M_{i,f}$,
the path integral (as transition amplitude) $\langle A|_{\Sigma_{f}}|T_{M}|A|_{\Sigma_{i}}\rangle$
is the diffeomorphism class of the smooth manifold relative to the
boundary. The diffeomorphism class of the boundary, however, is unique and the value of the path integral depends on the topology of the boundary
as well on the diffeomorphism class of the interior of $M_{i,f}$.
Therefore we will shortly write
\[
\langle\Sigma_{f}|T_{M}|\Sigma_{i}\rangle=\langle A|_{\Sigma_{f}}|T_{M}|A|_{\Sigma_{i}}\rangle
\]
and consider the sum of manifolds like $M_{i,h}=M_{i,f}\cup_{\Sigma_{f}}M_{f,h}$
with the amplitudes
\begin{equation}
\langle\Sigma_{h}|T_{M}|\Sigma_{i}\rangle=\sum_{A|\Sigma_{f}}\langle\Sigma_{h}|T_{M}|\Sigma_{f}\rangle\langle\Sigma_{f}|T_{M}|\Sigma_{i}\rangle\label{eq:decomposition-amplitudes}
\end{equation}
where we sum (or integrate) over the connections and frames on $\Sigma_{h}$
(see \cite{hawkingpathintegral1979}). Then the boundary term
\begin{equation}
S_{\partial}[\Sigma_{f}]=\intop_{\Sigma_{f}}\epsilon_{IJKL}(e^{I}\wedge e^{J}\wedge A^{KL})=\intop_{\Sigma_{f}}H\sqrt{h}d^{3}x\label{eq:boundary-term-EH}
\end{equation}
is needed where $H$ is the mean curvature of $\Sigma_{f}$ corresponding
to the metric $h$ at $\Sigma_{f}$ (as restriction of the 4-metric).
In the path integral (\ref{eq:path-integral}), one integrates over
the frames and connections. The possibility of singular frames was
discussed at some places (see \cite{Wit:89.2,Wit:89.3}). The cotetrad
field $e^{I}=e_{a}^{I}dx^{a}$ changes w.r.t. the smooth map $f:M\to M$
by $e_{a}^{I}(x)\, dx^{a}\mapsto e_{a}^{I}(x')\, dx'^{a}=e_{a}^{I}(f(x))(\partial_{b}f^{a}(x))dx^{b}$.
The transformation matrix $(\partial_{b}f^{a}(x))$ has maximal rank
$4$ for every regular value of the smooth map, but at the critical
points $x_{c}$ of $f$, some derivatives vanish and one has a smaller
rank at the point $x_{c}$, called a singular point. Then there is
no inverse frame (or tetrad field) at this point. Usually singular
frames are of this nature and one can decompose every singular frame
into a product of a regular frame and a (singular) transformation
induced by a smooth map. How can one interpret these singularities?
At this point one needs some differential topology. A homeomorphism
can be arbitrarily and accurately approximated by smooth mappings
(see \cite{Hir:76}, Theorem 2.6), i.e. in a neighborhood of a homeomorphism
one always finds a smooth map. Secondly, there is a special class
of smooth maps, the stable maps. Here, two smooth maps are stable
equivalent if both maps agree after a diffeomorphism of the corresponding
manifolds \cite{GolGui:73}. Here we are interested into smooth mappings
from 4-manifolds into 4-manifolds. By a deep result of Mather \cite{Mat:71},
stable mappings for this dimension are dense in all smooth mappings
of 4-manifolds. In \cite{AsselmeyerBrans2007}, we analyzed this situation:
the approximation of a homeomorphism by a stable map. If this smooth
map has no singularities then we can perturb them to a diffeomorphism.
For a singular map, however, we showed that it induces a change of the
smoothness structure. Then, a singular frame corresponds to a
regular frame in a different smoothness structure. The path integral
changed the domain of integration: 
\[
\intop_{\mbox{regular+singular frames}}De\quad\to\intop_{\mbox{smoothness structures}}De
\]
We remark that this change is unique for dimension four. No other
dimension has this plethora of smoothness structures which can be
used to express the singular frames. 

The inclusion of exotic smoothness changed the description of trivial
spaces like $\mathbb{R}^{4}$ completely. Instead of a single chart,
we have now an infinite sequence of charts or an infinite sequence
of 4-dimensional submanifolds. We will describe it more completely
later. Each submanifold is bounded by a 3-manifold (different from
a 3-sphere) and we obtain a sequence of 3-manifolds $Y_{0}\to Y_{1}\to Y_{2}\to\ldots$
characterizing the smoothness structure. The sequence of 3-manifolds
divides the path integral into a product 
\[
\langle Y_{0}|T_{M}|Y_{1}\rangle\langle Y_{1}|T_{M}|Y_{2}\rangle\langle Y_{2}|T_{M}|Y_{3}\rangle\cdots
\]
and we have to think about the boundary term (\ref{eq:boundary-term-EH}).
In \cite{TAMBrans2015,AsselmeyerRose:2012} we analyzed this
term: the boundary $Y_{n}$ seen as embedding into the spacetime $M$
can be described locally as spinor $\psi$ and one obtains for the
boundary term 
\begin{equation}
\intop_{Y_{n}}H\sqrt{h}d^{3}x=\intop_{Y_{n}}\bar{\psi}D\psi\sqrt{h}d^{3}x\label{eq:boundary-term-Dirac}
\end{equation}
the Dirac action with the Dirac operator $D$ and $|\psi|^{2}=const.$(see
\cite{Friedrich1998} for the construction of $\psi$). In particular
we obtained the eigenvalue equation $D\psi=H\psi$, i.e. the mean
curvature is the eigenvalue of the Dirac operator which has compact
spectrum (from the compactness of $Y_{n}$) or we obtained discrete
levels of geometry. This result enforced us to identify the 3-manifolds
(or the parts) with the matter content. Furthermore the path integral
of the boundary can be carried out by an integration along $\psi$
(see \cite{TAMKrol2014}).With some effort \cite{TAMBrans2015,AsselmeyerRose:2012},
one can extend this boundary term to a tubular neighborhood $Y_{n}\times[0,1]$
of the boundary $Y_{n}$. However, the relation (\ref{eq:boundary-term-Dirac})
is only true for simple (i.e. irreducible) 3-manifolds, i.e. for complements
of a knot admitting hyperbolic structure. For more complex 3-manifolds,
we have the following simple scheme: the knot complements are connected
by torus bundles (locally written as $T^{2}\times[0,1]$). Therefore
we also have to describe these bundles by using the boundary term.
In \cite{AsselmeyerRose:2012} we described this situation by using
the geometrical properties of these bundles and we will give a short
account of these ideas in subsection \ref{sub:The-Graviton-propagator}.
Simply expressed, in this bundle one has a flow of constant curvature
along the tube. The constant curvature connections are given by varying
the Chern-Simons functional. Now following Floer \cite{Flo:88}, the
4-dimensional version of this flow equation is the instanton equation
(or the self-dual equation) leading to the correct Yang-Mills functional
(Chern-Simons gives the Pontrjagin class and the instanton equation
makes it to the Yang-Mills functional). More importantly, the
three possible types of torus bundles fit very good into the current
scheme of three gauge field interactions (see \cite{AsselmeyerRose:2012} (section 8)).

\section{The Action Induced by Topology\label{sec:action-induced-by-top}}

Now we have the following picture: fermions as hyperbolic knot complements
and gauge bosons as torus bundles. Both components together are forming
an irreducible 3-manifold which is connecting to the remaining space
by a $S^{2}-$boundary (see the prime decomposition in Appendix B).
This connection via $S^{2}\times[0,1]$ ($S^{2}-$bundle) is the only
connection between matter and space. Here, there is only one interpretation:
this $S^{2}-$bundle must be interpreted as gravity. In this section
we will support this conjecture and construct the corresponding action.
At first we will fix the model, i.e. let $\Sigma_{M}$ and $\Sigma_{S}$
be the 3-manifolds for matter and space, respectively. The connected
sum $\#$ of both components represents the whole spatial component
\[
\Sigma=\Sigma_{M}\#\Sigma_{S}=\Sigma_{M}\cup_{S^{2}}\left(S^{2}\times[0,1]\right)\cup_{S^{2}}\Sigma_{S}=\Sigma_{M}\#S^{3}\#\Sigma_{S}
\]
of the spacetime. The decomposition above showed the geometry of the$S^{2}-$bundle
(in the sense of Thurston, see Appendix B) to be the spherical geometry
with isometry group $SO(3)$. The idea of the following construction
can be simply expressed: the 2-sphere $S^{2}$ explores locally the
curvature of the space where the curvature is given by the inverse
volume $\frac{1}{vol(S^{2})}$ of the 2-sphere $S^{2}$. The 2-sphere
can be written as a homogenous space $S^{2}=SO(3)/SO(2)$ also known
as Hopf bundle. As mentioned above, the geometry of the bundle $S^{2}\times[0,1]$
(interpreted as an equator region of $S^{3}$) is the spherical geometry
with isometry group $SO(3)$. So, as a local model we have an embedding
of a 3-manifold (as the spatial component for a fixed time) into the
spacetime with local Lorentz symmetry (represented by $SO(3,1)$).
From the mathematical point of view, it is a reductive Cartan geometry\cite{Wise2009,Wise2010}
over the homogenous space $SO(3,1)/SO(3)$, the 3-dimensional hyperbolic
space. For the moment, let us extend this symmetry to the spacetime
$M$ itself. A Cartan connection $A$ decomposes as a $so(3)-$valued
connection $\omega$ ($so(3)$ denotes the Lie algebra of $SO(3)$)
and a coframe field $e$ (with values in $so(3,1)/so(3)$) as
\[
A=\omega+\frac{1}{\ell}e
\]
by using the scale $\ell$(in agreement with the physical units) and
with curvature 
\begin{eqnarray*}
F & = & dA+A\wedge A\\
 & = & (d\omega+\omega\wedge\omega)+\frac{1}{\ell^{2}}e\wedge e=R+\frac{1}{\ell^{2}}e\wedge e
\end{eqnarray*}
Then for the spacetime (as 4-manifold), we interpret the Cartan connection
$A$ as the connection of the frame bundle (with respect to the Lorentz
structure). Now we have to think about what characterizes the $S^{2}-$bundle
in a 4-manifold, i.e. a surface bundle over a surface (at least locally).
It is known that a surface bundle over a surface is topologically
described by the Euler class as well as the Pontrjagin class (via the
Hirzebruch signature theorem). Therefore we choose the sum of the
Euler and Pontrjagin class for the frame bundle as action
\[
S=\intop_{M}\left(\epsilon^{ABCD}F_{AB}\wedge F_{CD}+\gamma F\wedge F\right)
\]
where the Pontrjagin class is weighted by a parameter $\gamma$. Using
the rules above, we obtain
\begin{eqnarray*}
S & = & \frac{1}{\ell^{2}}\intop_{M}\left(2\epsilon^{ABCD}e_{A}\wedge e_{B}\wedge R_{CD}+2\gamma e\wedge e\wedge R+\frac{(1+\gamma)}{\ell^{2}}e\wedge e\wedge e\wedge e\right)+\\
 &  & +\intop_{M}\left(\epsilon^{ABCD}R_{AB}\wedge R_{CD}+\gamma R\wedge R\right)\;,
\end{eqnarray*}
the Einstein-Hilbert action with cosmological constant and the Holst
action with Immirizo parameter as well the Euler and Pontrjagin class
for the reduced bundles. In this model, the curvature is changed locally
by adding a $S^{2}-$bundle. Then the scale $\ell^{2}$ has to agree
with the volume of the $S^{2}$. In the action we have the coupling
constant $\frac{1}{\ell^{2}}$ which has to agree with $1/L_{P}^{2}$
($L_{P}$ Planck length) to get in contact with Einsteins theory,
i.e. we must set
\[
\ell=L_{P}
\]
The agreement with the Einstein-Hilbert action showed that this approach
can describe gravity but it does not describe the global geometry.
Later we can show, however, that it must be the de Sitter space $SO(4,1)/SO(3,1)$ globally.

\section{Wild Embeddings: Geometric Expression for the Quantum State\label{sec:Wild-embeddings}}

In this section we will support our main hypothesis that an exotic
$\mathbb{R}^{4}$ has automatically a quantum geometry, but as noted
in the introduction, we must implicitly assume that the quantum-geometrical
state is realized in the exotic $\mathbb{R}^{4}$. Interestingly,
it follows from the physically motivated existence of a Lorentz
metric which is induced by a codimension-one foliation. Therefore
we will construct the foliation and the corresponding leaf space as
the space of observables (using ideas of Connes). This leaf space
is a non-commutative $C^{*}-$algebra with observable algebra a factor
$I\! I\! I_{1}$ von Neumann algebra. A state in this algebra can
be interpreted as a wild embedding which is also motivated by the
exotic smoothness structure. The classical state is the tame embedding.
Then, the deformation quantization of this tame embedding is the wild
embedding (see \cite{TAMKrol2013}). In principle, the wild
embedding determines the $C^{*}-$algebra completely. This algebra
is generated by holonomies along connections of constant curvature.
It is known from mathematics that this algebra (forming a so-called
character variety \cite{CulSha:83}) determines the geometrical structure
of the 3-manifold (along the way of Thurston \cite{Scott1983,Thu:97}).
The main structure in this approach is the fundamental group, i.e.
the group of closed, non-contractible curves in a manifold. The quantization
of this group (as an expression of the classical geometry) gives the
so-called skein algebra of knots in this manifold. We will relate
this skein algebra to the leaf space above. On the way to show this
relation, we will obtain the generator of the translation from one
3-manifold into another 3-manifold, i.e. the time together with the
Hamiltonian.

\subsection{Exotic $\mathbb{R}^{4}$ and its Foliation}

In section \ref{sec:pathint}, we described the sequence of 3-manifolds 
\[
Y_{1}\to Y_{2}\to\cdots\to Y_{\infty}=Y_{r}
\]
characterizing the exotic smoothness structure of $R^{4}$. Then $0-$framed
surgery along this pretzel knot produces $Y_{1}$ whereas the $n$-th
untwisted Whitehead double will give $Y_{n}$. For large $n$, the
structure of the Casson handle is contained in the topology of $Y_{n}$
and in the limit $n\to\infty$ we obtain $Y_{r}$ (which is now a
wild embedding $Y_{r}\subset\mathbf{R}^{4}$ in the standard $\mathbf{R}^{4}$
given by the embedding of the small exotic $R^{4}$, see above). What do we know about the structure of $Y_{n}$ or $Y_{r}$ in general?
The compact subset $K$ is a 4-manifold constructed by a pair of one
1-handle and one 2-handle which topologically cancel. The boundary
of $K$ is a compact 3-manifold having the first Betti number $b_{1}=1.$
This information is also contained in $Y_{r}$. By the work of Freedman
\cite{Fre:82}, every Casson handle is topologically $D^{2}\times\mathbb{R}^{2}$
(relative to the attaching region) and therefore $Y_{r}$ must be
the boundary of $D^{4}$ (the Casson handle trivializes $K$ to be
$D^{4}$), i.e.\emph{ $Y_{r}$ is a wild embedded 3-sphere $S^{3}$}.
Then we obtain two different descriptions of $R^{4}$:
\begin{enumerate}
\item as a sequence of 3-manifolds $Y_{n}$ (all having the first Betti
number $b_{1}=1$) as boundaries of the neighborhood of $K$ with
increasing size and
\item as a global hyperbolic space of $R^{4}\setminus K$ written as $S_{\infty}^{3}\times\mathbb{R}$
where $S_{\infty}^{3}$ is a wild embedded 3-sphere (which looks differently
for different $t\in\mathbb{R}$).
\end{enumerate}
The first description gives a non-trivial but smooth foliation but
there is no global spatial space. In contrast to this highly non-trivial
foliation, the second description gives a global foliated spacetime
containing a global spatial component, the wild embedded 3-sphere.
In the first description we have a complex, relational description
with no global time-like slices. Here, there only is a local coordinate system (with its own eigenzeit). This relational view has the big
advantage that the simplest parts are also simple submanifolds (only
finite surfaces with boundary). In contrast, the second description
introduces a global foliation into equal time slices. Then the complexity
is contained into the spatial component which is now a wild embedding
(i.e. a space with an infinite number of polygons). This second approach
will be described in the next subsection. So, lets concentrate on
the first approach. Every 3-manifold $Y_{n}$ admits a codimension-one
($PSL(2,\mathbb{R})-$invariant) foliation (see \cite{TAMKrol2009}
for the details). By the description of the exotic $R^{4}$ using
the sequence of 3-manifolds
\[
Y_{1}\to Y_{2}\to Y_{3}\to\cdots
\]
we also get a foliation of the exotic $R^{4}$. The foliation on $Y_{n}$
is defined by a $PSL(2,\mathbb{R})-$invariant one-form $\omega$
which is integrable $d\omega\wedge\omega=0$ and defines another one-form
$\eta$ by $d\omega=-\eta\wedge\omega$. Then the integral
\[
GV(Y_{n})=\intop_{Y_{n}}\eta\wedge d\eta
\]
is known as Godbillon-Vey number $GV(Y_{n})$ with the class $gv=\eta\wedge d\eta$.
From the physics point of view, it is the abelian Chern-Simons functional.
The Godbillon-Vey class characterizes the codimension-one foliation
for the 3-manifold $Y_{n}$ (see the Appendix B for more details).
The foliation is very complicated. In \cite{ReinhartWood1973} the local
structure was analyzed. Let $\kappa,\tau$ be the curvature and torsion
of a normal curve, respectively. Furthermore, let $T,N,Z$ be the
frame formed by this vector field dual to the one-forms $\omega,\eta,\xi$
and let $l_{T}$ be the second fundamental form of leaf. Then the
Godbillon-Vey class is locally given by 
\[
\eta\wedge d\eta=\kappa^{2}\left(\tau+l_{T}(N,Z)\right)\omega\wedge\eta\wedge\xi
\]
where $\tau\not=0$ for $PSL(2,\mathbb{R})$ invariant foliations
i.e. $[Z,N]=Z$, $[N,T]=T$ and $[Z,T]=N$. Recall that a foliation
$(M,F)$ of a manifold $M$ is an integrable subbundle $F\subset TM$
of the tangent bundle $TM$. The leaves $L$ of the foliation $(M,F)$
are the maximal connected submanifolds $L\subset M$ with $T_{x}L=F_{x}\:\forall x\in L$.
We denote with $M/F$ the set of leaves or the leaf space. Now one
can associate to the leaf space $M/F$ a $C^{*}$-algebra $C(M,F)$
by using the smooth holonomy groupoid $G$ of the foliation (see Connes
\cite{Connes1984}). According to Connes \cite{Connes94}, one assigns
to each leaf $\ell\in X$ the canonical Hilbert space of square-integrable
half-densities $L^{2}(\ell)$. This assignment, i.e. a measurable
map, is called a random operator forming a von Neumann $W(M,F)$.
A deep theorem of Hurder and Katok \cite{HurKat:84} for foliations
with non-zero Godbillon-Vey invariant states that this foliation has
to contain a factor $I\! I\! I$ von Neumann algebra. As shown in
\cite{TAMKrol2011a}, the von Neumann algebra for the foliation
of $Y_{n}$ and for the exotic $R^{4}$ is a factor $I\! I\! I_{1}-$algebra.
For the construction of this algebra, one needs the concept of a holonomy
groupoid. Foliations are determined by the holonomies of closed curves
in a leaf and the transport of this closed curve together with the
holonomy from the given leaf to another leaf. Now one may ask why
one considers only closed curves. Let $PM$ the space of all paths
in a manifold then this space admits a fibration over the space of
closed paths $\Omega M$ (also called loop space) with fiber the constant
paths (therefore homeomorphic to $M$), see \cite{BoTu:95}. Then,
a curve is determined up to deformation (i.e. homotopy) by a closed
path. Consider now a closed curve $\gamma$ in a leaf $\ell$ and
let act a diffeomorphism on $\ell$. Then the curve $\gamma$ is modified
as well to $\gamma'$ but $\gamma$ and $\gamma'$ are related by
a (smooth) homotopy. Therefore to guarantee diffeomorphism invariance
in this approach, one has to consider all closed curves up to homotopy.
This structure can be made into a group (using concatenation of paths
as group operation) called the fundamental group $\pi_{1}(\ell)$
of the leaf. Above we spoke about holonomy but a holonomy needs a
connection of some bundle which we did not introduce until now. But
Connes \cite{Connes1984} described a way to circumvent this difficulty:
Given a leaf $\ell$ of $(M,F)$ and two points $x,y\in\ell$ of this
leaf, any simple path $\gamma$ from $x$ to $y$ on the leaf $\ell$
uniquely determines a germ $h(\gamma)$ of a diffeomorphism from a
transverse neighborhood of $x$ to a transverse neighborhood of $y$.
The germ of diffeomorphism $h(\gamma)$ only depends upon the homotopy
class of $\gamma$ in the fundamental group of the leaf $\ell$, and
is called the holonomy of the path $\gamma$. All fundamental groups
of all leafs form the fundamental groupoid. The holonomy groupoid
of a leaf $\ell$ is the quotient of its fundamental groupoid by the
equivalence relation which identifies two paths $\gamma$ and $\gamma'$
from $x$ to $y$ (both in $\ell$) iff $h(\gamma)=h(\gamma')$. Then
the von Neumann algebra of the foliation is the convolution algebra
of the holonomy groupoid which will be constructed later for the wild
embedding.

\subsubsection{Intermezzo: Factor $I\! I\! I$ and Tomita-Takesaki Modular Theory}

Remember a von Neumann algebra is an involutive subalgebra $M$ of
the algebra of operators on a Hilbert space $H$ that has the property
of being the commutant of its commutant: $(M')'=M$. This property
is equivalent to saying that $M$ is an involutive algebra of operators
that is closed under weak limits. A von Neumann algebra $M$ is said
to be hyperfinite if it is generated by an increasing sequence of
finite-dimensional subalgebras. Furthermore we call $M$ a factor
if its center is equal to $\mathbb{C}$. It is a deep result of Murray
and von Neumann that every factor $M$ can be decomposed into 3 types
of factors $M=M_{I}\oplus M_{II}\oplus M_{III}$. The factor $I$
case divides into the two classes $I_{n}$ and $I_{\infty}$ with
the hyperfinite factors $I_{n}=M_{n}(\mathbb{C})$ the complex square
matrices and $I_{\infty}=\mathcal{L}(H)$ the algebra of all operators
on an infinite-dimensional Hilbert space $H$. The hyperfinite $I\! I$
factors are given by $I\! I_{1}=Cliff_{\mathbb{C}}(E)$, the Clifford
algebra of an infinite-dimensional Euclidean space $E$, and $I\! I_{\infty}=I\! I_{1}\otimes I_{\infty}$.
The case $I\! I\! I$ remained mysterious for a long time. Now we
know that there are three cases parametrized by a real number $\lambda\in[0,1]$:
$I\! I\! I_{0}=R_{W}$ the Krieger factor induced by an ergodic flow
$W$, $I\! I\! I_{\lambda}=R_{\lambda}$ the Powers factor for $\lambda\in(0,1)$
and $I\! I\! I_{1}=R_{\infty}=R_{\lambda_{1}}\otimes R_{\lambda_{2}}$
the Araki-Woods factor for all $\lambda_{1},\lambda_{2}$ with $\lambda_{1}/\lambda_{2}\notin\mathbb{Q}$.
We remark that all factor  $I\! I\! I$ cases are induced by infinite
tensor products of the other factors. One example of such an infinite
tensor space is the Fock space in quantum field theory. 

The modular theory of von Neumann algebras (see also \cite{Borchers2000})
has been discovered by M.~Tomita~\cite{Tomita1967} in 1967 and put on solid grounds by
M.~Takesaki~\cite{Takesaki1970} around 1970. It is a very deep theory that,
to every von Neumann algebra $\mathcal{M}\subset\mathcal{B}(\mathcal{H})$
acting on a Hilbert space $\mathcal{H}$, and to every vector $\xi\in\mathcal{H}$
that is {cyclic}, i.e.~$\overline{(\mathcal{M}\xi)}=\mathcal{H}$,
and {separating}, i.e.~for $A\in\mathcal{M}$, $A\xi=0\rightarrow A=0$,
associates: 
\begin{itemize}
\item a one-parameter unitary group $t\mapsto\Delta^{it}\in\mathcal{B}(\mathcal{H})$ 
\item and a conjugate-linear isometry $J:\mathcal{H}\to\mathcal{H}$ such
that:{\samepage 
\begin{gather*}
\Delta^{it}\mathcal{M}\Delta^{-it}=\mathcal{M},\quad\forall\, t\in\mathbb{R},\qquad\text{and}\qquad J\mathcal{M}J=\mathcal{M}',
\end{gather*}
where the {commutant} $\mathcal{M}'$ of $\mathcal{M}$ is defined
by $\mathcal{M}'\!:=\!\{A'\in\mathcal{B}(\mathcal{H})\,|\,[A',A]_{-}\!=0,\forall\, A\!\in\!\mathcal{B}(\mathcal{H})\}$.} 
\end{itemize}
More generally, given a von Neumann algebra $\mathcal{M}$ and a faithful
normal state%
\footnote{$\omega$ is faithful if $\omega(x)=0\rightarrow x=0$; it is normal
if for every increasing bounded net of positive elements $x_{\lambda}\to x$,
we have $\omega(x_{\lambda})\to\omega(x)$.%
} (more generally for a faithful normal semi-finite weight) $\omega$
on the algebra $\mathcal{M}$, the modular theory allows to create
a one-parameter group of $*$-automorphisms of the algebra $\mathcal{M}$,
\[
\sigma^{\omega}:t\mapsto\sigma_{t}^{\omega}\in\text{Aut}(\mathcal{M}),\qquad\text{with}\quad t\in\mathbb{R},
\]
such that: 
\begin{itemize}
\item in the Gel'fand--Na\u{\i}mark--Segal representation $\pi_{\omega}$
induced by the weight $\omega$, on the Hilbert space $\mathcal{H}_{\omega}$,
the {modular automorphism group} $\sigma^{\omega}$ is implemented
by a unitary one-parameter group $t\mapsto\Delta_{\omega}^{it}\in\mathcal{B}(\mathcal{H}_{\omega})$
i.e.~we have $\pi_{\omega}(\sigma_{t}^{\omega}(x))=\Delta_{\omega}^{it}\pi_{\omega}(x)\Delta_{\omega}^{-it}$,
for all $x\in\mathcal{M}$ and for all $t\in\mathbb{R}$; 
\item there is a conjugate-linear isometry $J_{\omega}:\mathcal{H}_{\omega}\to\mathcal{H}_{\omega}$,
whose adjoint action implements a~{modular anti-isomorphism}
$\gamma_{\omega}:\pi_{\omega}(\mathcal{M})\to\pi_{\omega}(\mathcal{M})'$,
between $\pi_{\omega}(\mathcal{M})$ and its commutant $\pi_{\omega}(\mathcal{M})'$,
i.e.~for all $x\in\mathcal{M}$, we have $\gamma_{\omega}(\pi_{\omega}(x))=J_{\omega}\pi_{\omega}(x)J_{\omega}$. 
\end{itemize}
The operators $J_{\omega}$ and $\Delta_{\omega}$ are called respectively
the {modular conjugation operator} and the {modular operator}
induced by the state (weight) $\omega$. We will call ``{modular
generator}'' the self-adjoint generator of the unitary one-parameter
group $t\mapsto\Delta_{\omega}^{it}$ as defined by Stone's theorem
i.e.~the operator 
\begin{equation}
K_{\omega}:=\log\Delta_{\omega},\qquad\text{so that}\quad\Delta_{\omega}^{it}=e^{iK_{\omega}t}.\label{eq:modular-generator}
\end{equation}

The modular automorphism group $\sigma^{\omega}$ associated to $\omega$
is the unique one-parameter automorphism group that satisf\/ies the
Kubo--Martin--Schwinger {(KMS-condition)} with respect to the state
(or more generally a normal semi-finite faithful weight) $\omega$,
at inverse temperature $\beta=-1$, i.e. 
\[
\omega(\sigma_{t}^{\omega}(x))=\omega(x),\qquad\forall\, x\in\mathcal{M}
\]
and for all $x,y\in\mathcal{M}$. 

Using Tomita-Takesaki-theory, one has a continuous decomposition (as
crossed product) of any factor $I\! I\! I$ algebra $M$ into a factor
$I\! I_{\infty}$ algebra $N$ together with a one-parameter group%
\footnote{The group $\mathbb{R}_{+}^{*}$ is the group of positive real numbers
with multiplication as group operation also known as Pontrjagin dual.%
} $\left(\theta_{\lambda}\right)_{\lambda\in\mathbb{R}_{+}^{*}}$ of
automorphisms $\theta_{\lambda}\in Aut(N)$ of $N$, i.e. one obtains

\begin{equation}
M=N\rtimes_{\theta}\mathbb{R}_{+}^{*}\quad.\label{eq:crossed-product-factor-III}
\end{equation}

That means, there is a foliation induced from the foliation producing
this $I\! I_{\infty}$ factor. Connes \cite{Connes94} (in section
I.4 page 57ff) constructed the foliation $F'$ canonically associated
to the foliation $F$ of factor $I\! I\! I_{1}$ above having the
factor $I\! I_{\infty}$ as von Neumann algebra. In our case it is
the horocycle flow: Let $P$ the polygon on the hyperbolic space $\mathbb{H}^{2}$
determining the foliation above. $P$ is equipped with the hyperbolic
metric $2|dz|/(1-|z|^{2})$ together with the collection $T_{1}P$
of unit tangent vectors to $P$. A horocycle in $P$ is a circle contained
in $P$ which touches $\partial P$ at one point, but from the classification
of factors, we know that $I\! I_{\infty}$ is also splitted into
\[
I\! I_{\infty}=I\! I_{1}\otimes I_{\infty}
\]
so that every factor $I\! I\! I$ is determined by the factor $I\! I_{1}$.
The factor $I_{\infty}$ are the compact operators in the Hilbert
space. With an important observation we will close this intermezzo.
The factor $I\! I_{\infty}$ admits an action of the group $\mathbb{R}_{+}^{*}$
by automorphisms so that the crossed product (\ref{eq:crossed-product-factor-III})
is the factor $I\! I\! I_{1}$. The corresponding invariant, the flow
of weights $\bmod(M)$, was determined by Connes \cite{Connes94}
to be the Godbillon-Vey invariant. Therefore \emph{the modular generator
above is given by the Godbillon-Vey invariant, i.e. this invariant
is the Hamiltonian of the theory.}

\subsubsection{Construction of a State\label{sub:Construction-of-state}}

Then the $C^{*}$-algebra $C_{r}^{*}(M,F)$ of the foliation $(M,F)$
is the $C^{*}$-algebra $C_{r}^{*}(G)$ of the smooth holonomy groupoid
$G$. For completeness we will present the explicit construction (see
\cite{Connes94} sec. II.8). The basic elements of $C_{r}^{*}(M,F)$)
are smooth half-densities with compact supports on $G$, $f\in C_{c}^{\infty}(G,\Omega^{1/2})$,
where $\Omega_{\gamma}^{1/2}$ for $\gamma\in G$ is the one-dimensional
complex vector space $\Omega_{x}^{1/2}\otimes\Omega_{y}^{1/2}$, where
$s(\gamma)=x,t(\gamma)=y$, and $\Omega_{x}^{1/2}$ is the one-dimensional
complex vector space of maps from the exterior power $\Lambda^{k}F_{x}$
,$k=\dim F$, to $\mathbb{C}$ such that 
\[
\rho(\lambda\nu)=|\lambda|^{1/2}\rho(\nu)\qquad\forall\nu\in\Lambda^{k}F_{x},\lambda\in\mathbb{R}\:.
\]
For $f,g\in C_{c}^{\infty}(G,\Omega^{1/2})$, the convolution product
$f*g$ is given by the equality
\[
(f*g)(\gamma)=\intop_{\gamma_{1}\circ\gamma_{2}=\gamma}f(\gamma_{1})g(\gamma_{2})
\]
Then we define via $f^{*}(\gamma)=\overline{f(\gamma^{-1})}$ a $*$-operation
making $C_{c}^{\infty}(G,\Omega^{1/2})$ into a $*$-algebra. For each
leaf $L$ of $(M,F)$ one has a natural representation of $C_{c}^{\infty}(G,\Omega^{1/2})$
on the $L^{2}$ space of the holonomy covering $\tilde{L}$ of $L$.
Fixing a base point $x\in L$, one identifies $\tilde{L}$ with $G_{x}=\left\{ \gamma\in G,\, s(\gamma)=x\right\} $and
defines the representation
\[
(\pi_{x}(f)\xi)(\gamma)=\intop_{\gamma_{1}\circ\gamma_{2}=\gamma}f(\gamma_{1})\xi(\gamma_{2})\qquad\forall\xi\in L^{2}(G_{x}).
\]
The completion of $C_{c}^{\infty}(G,\Omega^{1/2})$ with respect to
the norm 
\[
||f||=\sup_{x\in M}||\pi_{x}(f)||
\]
makes it into a $C^{*}$-algebra $C_{r}^{*}(M,F)$. Among all elements
of the $C^{*}$-algebra, there are distinguished elements, idempotent
operators or projectors having a geometric interpretation in the foliation.
For later use, we will construct them explicitly (we follow \cite{Connes94}
sec. II.8.$\!\beta$ closely). Let $N\subset M$ be a compact submanifold
which is everywhere transverse to the foliation (thus $\dim(N)=\mathrm{codim}(F)$).
A small tubular neighborhood $N'$ of $N$ in $M$ defines an induced
foliation $F'$ of $N'$ over $N$ with fibers $\mathbb{R}^{k},\, k=\dim F$.
The corresponding $C^{*}$-algebra $C_{r}^{*}(N',F')$ is isomorphic
to $C(N)\otimes\mathcal{K}$ with $\mathcal{K}$ the $C^{*}$-algebra
of compact operators. In particular it contains an idempotent $e=e^{2}=e^{*}$,
$e=1_{N}\otimes f\in C(N)\otimes\mathcal{K}$ , where $f$ is a minimal
projection in $\mathcal{K}$. The inclusion $C_{r}^{*}(N',F')\subset C_{r}^{*}(M,F)$
induces an idempotent in $C_{r}^{*}(M,F)$ which is given by a closed
curve in $M$ transversal to the foliation.

In case of the foliation above (of the 3-manifolds $Y_{n}$), one
has the foliation of the polygon $P$ in $\mathbb{H}^{2}$ and a circle
$S^{1}$ attached to every leaf of this foliation. Therefore we have
the leafs $S^{1}\times[0,1]$ and the $S^{1}$is the closed curve
transversal to the foliation. Then every leaf defines (using the
isomorphism $\pi_{1}(S^{1}\times[0,1])=\pi_{1}(S^{1})=\mathbb{Z}$)
an idempotent represented by the fiber $S^{1}$ forming a base for
the GNS representation of the $C^{*}-$algebra. Now we are able to
construct a state in this algebra.

\begin{figure}
\includegraphics[scale=0.5]{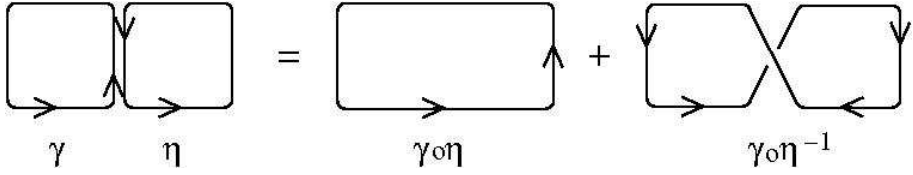}
\caption{Mandelstam identity as skein relation (see Fig. \ref{fig:skein-crossings} and subsection \ref{sub:Drinfeld-Turaev-Quantization}).\label{fig:Mandelstam-identity-as-skein}}
\end{figure}

A state is a linear functional $\omega:C_{r}^{*}(M,F)\to\mathbb{C}$
so that $\omega(x\cdot x^{*})\geq0$ and $\omega(\mathbb{I}_{C_{r}^{*}(M,F)})=1$.
Elements of $C_{r}^{*}(M,F)$ are half-densities with a support along
some closed curve (as part of the holonomy groupoid). In a first step,
one can use the GNS-representation of the $C*-$algebra $C_{r}^{*}(M,F)$
by a map $C_{r}^{*}(M,F)\to\mathcal{B}(H)$ in to the bounded operators
of a Hilbert space. By the theorem of Fr{\'e}chet-Riesz, every linear
functional can be represented by the scalar product of the Hilbert
space for some vector. To determine the linear functionals, we have
to investigate the geometry of the foliation. The foliation was constructed
to be $PSL(2,\mathbb{R})-$invariant, i.e. fixing the upper half space
$\mathbb{H}^{2}$. Then we considered the unit tangent vectors of
the tangent bundle over $\mathbb{H}^{2}$ defining the $\tilde{SL}(2,\mathbb{R})-$geometry.
But more is true. Every part of the 3-manifold $Y_{n}$ is a knot/link
complement with hyperbolic structure with isometry group $PSL(2,\mathbb{C})$
where the other geometric structures like $\tilde{SL}(2,\mathbb{R})$
and $PSL(2,\mathbb{R})$ embed. Here we remark the known fact that
every $PSL(2,\mathbb{C})$-geometry lifts uniquely to $SL(2,\mathbb{C})$
(the double cover). Therefore, to model the holonomy, we have to choose
a flat $SL(2,\mathbb{C})-$connection and write it as the well-known
integral of the connection 1-form along a closed curve. The linear
functional is the trace of this integral (seen as matrix using a representation
of $SL(2,\mathbb{C})$) known as Wilson loop. One can use the well-known
identity 
\[
Tr(A)\cdot Tr(B)=Tr(AB)+Tr(AB^{-1})
\]
for $SL(2,\mathbb{C})$ which goes over to the Wilson loops. Let $W_{\gamma}[A]$
be the Wilson loop of a connection $A$ along the closed curve $\gamma$.
Then the relation of the Wilson loops
\[
W_{\gamma}[A]\cdot W_{\eta}[A]=W_{\gamma\circ\eta}[A]+W_{\gamma\circ\eta^{-1}}[A]
\]
for two intersecting curves $\gamma$ and $\eta$ is known as the
Mandelstam identity for intersecting loops, see Fig. \ref{fig:Mandelstam-identity-as-skein}
for a visualization. 

 This relation is also known from another area: knot theory. There,
it is the Kauffman bracket skein relation used to define the Kauffman
knot polynomial. Therefore we obtain a state in the $C*-$algebra
by a closed curve in the leaf which extends to a knot (an embedded,
closed curve) in a submanifold of the 3-manifold defined up to the
skein relation. Finally:
\[
\mbox{State }\omega\mbox{ over leaf }\ell\longleftrightarrow\mbox{element of Kauffman skein module for }\mbox{\ensuremath{\ell\times}[0,1]}
\]
We will later explain this correspondence as a deformation quantization.
We will close this subsection by some remarks. Every representation
$\pi_{1}(M)\to SL(2,\mathbb{C})$ defines (up to conjugacy) a flat
connection. At the same time it defines also a hyperbolic structure
on $Y_{n}$ (for $M=Y_{n}$). By the argumentation above, the quantized
version of this geometry (as defined by the $C*-$algebra of the foliation)
is given by the skein space (see subsection \ref{sub:Drinfeld-Turaev-Quantization} for the definition of the skein space).

\subsection{The Wild Embedded 3-Sphere = Quantum (Geometric) State}

Our previous work implied that the transition from the standard $\mathbf{R}^{4}$
to a small exotic $R^{4}$ has much to do with Quantum Gravity (QG).
Therefore one would expect that a submanifold in the standard $\mathbf{R}^{4}$
with an appropriated geometry represents a classical state. Before
we construct this state, there is a lot to say about the wild
embedded 3-sphere as a quantum state.

\subsubsection{The Wild Embedded 3-Sphere}
\begin{figure}
\centering \resizebox{0.7\textwidth}{!}{ \includegraphics{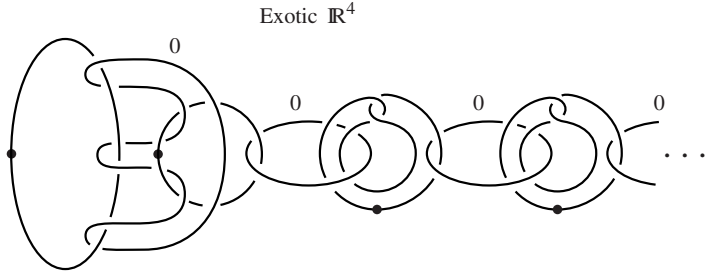}
} \caption{Handle picture of the small exotic $\mathbb{R}^{4}$, the components
with the dot are 1-handles and without the dot are 2-handles.\label{fig:Handle-picture-of-exotic-R4}}
\end{figure}

To describe this wild 3-sphere, we will construct the sequence of
$Y_{n}$ by using the example of \cite{Biz:95,BizGom1996} which was
already partly explained in subsection \ref{sub:Small-exotic-R4-sequnce-3MF}.
At first we remark that the interior of the handle body in Figure
\ref{fig:Handle-picture-of-exotic-R4} is the $R^{4}$. 

\begin{figure}
\centering \resizebox{0.6\textwidth}{!}{ \includegraphics{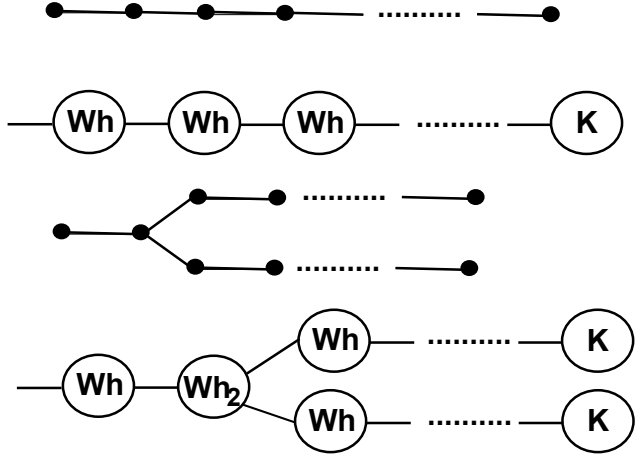}
} \caption{Schematic picture for the splitting of the knot complement $C(\mathcal{K}_{+})$
(above) and in the more general case $C(\mathcal{K_{T}})$ (below).\label{fig:Splitting-of-knot-complement} }
\end{figure}

The Casson handle for this $R^{4}$ is given by the simplest tree
$\mathcal{T}_{+}$, one positive self-intersection for each level.
The compact 4-manifold inside of $R^{4}$ can be seen in Figure \ref{fig:link-picture-for-K}
as a handle body. The 3-manifold $Y_{n}$ surrounding this compact submanifold
$K$ is given by surgery ($0-$framed) along the link in Figure \ref{fig:Handle-picture-of-exotic-R4}
with a Casson handle of $n-$levels. In \cite{Ganzel2006}, this case
is explicitly discussed. $Y_{n}$ is given by $0-$framed surgery
along the $n$-th untwisted Whitehead double of the pretzel $(-3,3-3)$
knot (see Figure \ref{fig:pretzel-knot}). Obviously, there is a sequence
of inclusions 
\[
\ldots\subset Y_{n-1}\subset Y_{n}\subset Y_{n+1}\subset\ldots\to Y_{\mathcal{T}_{+}}
\]
with the 3-manifold $Y_{\mathcal{T}_{+}}$ as limit. Let $\mathcal{K}_{+}$
be the corresponding (wild) knot, i.e. the $\infty$-th untwisted Whitehead
double of the pretzel knot $(-3,3,-3)$ (or the knot $9_{46}$ in
Rolfson notation). The surgery description of $Y_{\mathcal{T}_{+}}$induces
the decomposition 
\begin{equation}
Y_{\mathcal{T}_{+}}=C(\mathcal{K}_{+})\cup\left(D^{2}\times S^{1}\right)\qquad C(\mathcal{K}_{+})=S^{3}\setminus\left(\mathcal{K}_{+}\times D^{2}\right)\label{eq:surgery-description-of-Y}
\end{equation}
where $C(\mathcal{K}_{+})$ is the knot complement of $\mathcal{K}_{+}$.
In \cite{Budney2006}, the splitting of knot complements was described.
Let $K_{9_{46}}$ be the pretzel knot $(-3,3,-3)$ and let $L_{Wh}$
be the Whitehead link (with two components). Then the complement $C(K_{9_{46}})$
has one torus boundary whereas the complement $C(L_{Wh})$ has two
torus boundaries. Now according to \cite{Budney2006}, one obtains
the splitting 
\[
C(\mathcal{K}_{+})=C(L_{Wh})\cup_{T^{2}}\cdots\cup_{T^{2}}C(L_{Wh})\cup_{T^{2}}C(K_{9_{46}})
\]
and we will describe each part separately (see Figure \ref{fig:Splitting-of-knot-complement}).

At first the knot $K_{9_{46}}$ is a hyperbolic knot, i.e. the interior
of the 3-manifold $C(K_{9_{46}})$ admits a hyperbolic metric. By
the work of Gabai \cite{Gabai1983}, $C(K_{9_{46}})$admits a codimension-one
foliation. The Whitehead link is a hyperbolic link but we need more:
the Whitehead link is a fibered link of genus $1$. That is, there
is a fibration of the link complement $\pi:C(L_{Wh})\to S^{1}$ over
the circle so that $\pi^{-1}(p)$ is a surface of genus $1$ (Seifert
surface) for all $p\in S^{1}$. Now we will also describe the changes
for a general tree. At first we will modify the Whitehead link: we
duplicate the linked circle, i.e. there are as many circles as branching
in the tree to get the link $Wh_{n}$ with $n+1$ components. Then
the complement of $Wh_{n}$ has also $n+1$ torus boundaries and it
also fibers over $S^{1}$. With the help of $Wh_{n}$ we can build
every tree $\mathcal{T}$. Now the 3-manifold $Y_{\mathcal{T}}$ is
given by $0-$framed surgery along the $\infty$-th untwisted ramified
(usage of $Wh_{n}$) Whitehead double of a knot $k$, denoted by the
link $\mathcal{K_{T}}$. The tree $\mathcal{T}$ has one root, then
$Y_{\mathcal{T}}$ is given by 
\[
Y_{\mathcal{T}}=C(\mathcal{K_{T}})\cup\left(D^{2}\times S^{1}\right)
\]
and the complement $C(\mathcal{K_{T}})$ splits like the tree into
complements of $Wh_{n}$ and one copy of $C(k)$ (see Figure \ref{fig:Splitting-of-knot-complement}).
Using a deep result of Freedman \cite{Fre:82}, we obtain:\\
\emph{$Y_{\mathcal{T}}$ is a wild embedded 3-sphere $S_{\infty}^{3}$.}

\subsubsection{Reconstruction of the Spatial Space by Using a State}

Our result about the existence of a codimension-one foliation for
$Y_{\mathcal{T}}$ can be simply expressed: foliations are characterized
by the holonomy properties of the leafs. This principle is also the
corner stone for the usage of non-commutative geometry as description
of the leaf space. In the previous subsection, we already characterized
the state as an element of the Kauffman skein module. Here we are interested
in a reconstruction of the underlying space but now assuming a global
foliation so that we will obtain the whole spatial space. 

Starting point is the state constructed in the subsection \ref{sub:Construction-of-state}.
Here, we got a relation between the state $\omega$ as linear functional
over the algebra and the Kauffman skein module. Using this relation,
we consider a leaf $\ell=S^{1}\times[0,1]$ and the 3-dimensional
extension as solid torus $S^{1}\times D^{2}$. The Kauffman skein
module $K(S^{1}\times D^{2})$ is polynomial algebra with one generator
(the loop around $S^{1}$). Now we consider one 3-manifold $Y_{n}$
with the corresponding foliation. Using the splitting above, the Kauffman
skein module $K(Y_{n})$ is determined by the skein module for the
parts, i.e. by the knot complements. Therefore we have to consider
the skein module for hyperbolic 3-manifolds. Hyperbolic 3-manifolds
contain special surfaces, called essential or incompressible surfaces,
see Appendix C. It is known \cite{BullockFrohmanKania-Bartoszynska1999} that the skein module
of 3-manifolds containing essential surfaces is not finitely generated.
Therefore, the state itself is not finitely generated. If we use the
leaf $S^{1}\times D^{2}$ as a local model for one generator then
we will obtain an infinitely complicated 3-manifold made from pieces
$S^{1}\times D^{2}$ so that the corresponding generators are not
related to each other. An example of this structure is the Whitehead
manifold having a non-finitely generated Kauffman skein module \cite{Abchir1998}.
In general we will obtain a wild embedded 3-manifold by using this
simple pieces. By the argumentation in the previous subsection we
know that this wild embedded 3-manifold is the wild embedded 3-sphere
$Y_{\mathcal{T}}$. Finally we obtain:
\[
\mbox{State }\omega\longleftrightarrow\mbox{\mbox{wild embedded 3-sphere }}Y_{\mathcal{T}}
\]
the state $\omega$ is realized by some wild embedded 3-sphere.

\subsubsection{Construction of the Operator Algebra}

Following \cite{TAMKrol2013} we will construct a $C^{*}-$algebra
from the wild embedded 3-sphere. Let $I:S^{3}\to\mathbb{R}^{4}$ be
a wild embedding of codimension-one so that $I(S^{3})=S_{\infty}^{3}=Y_{\mathcal{T}}$.
Now we consider the complement $\mathbb{R}^{4}\setminus I(S^{3})$
which is non-trivial, i.e. $\pi_{1}(\mathbb{R}^{4}\setminus I(S^{3}))=\pi\not=1$.
Now we define the $C^{*}-$algebra $C^{*}(\mathcal{G},\pi$) associated
to the complement $\mathcal{G}=\mathbb{R}^{4}\setminus I(S^{3})$
with group $\pi=\pi_{1}(\mathcal{G})$. If $\pi$ is non-trivial then
this group is not finitely generated. From an abstract point of view,
we have a decomposition of $\mathcal{G}$ by an infinite union
\[
\mathcal{G}=\bigcup_{i=0}^{\infty}C_{i}
\]
of 'level sets' $C_{i}$. Then every element $\gamma\in\pi$ lies
(up to homotopy) in a finite union of levels. 

The basic elements of the $C^{*}-$algebra $C^{*}(\mathcal{G},\pi$)
are smooth half-densities with compact supports on $\mathcal{G}$,
$f\in C_{c}^{\infty}(\mathcal{G},\Omega^{1/2})$, where $\Omega_{\gamma}^{1/2}$
for $\gamma\in\pi$ is the one-dimensional complex vector space of
maps from the exterior power $\Lambda^{k}L$ ($\dim L=k$), of the
union of levels $L$ representing $\gamma$, to $\mathbb{C}$ such
that 
\[
\rho(\lambda\nu)=|\lambda|^{1/2}\rho(\nu)\qquad\forall\nu\in\Lambda^{2}L,\lambda\in\mathbb{R}\:.
\]
For $f,g\in C_{c}^{\infty}(\mathcal{G},\Omega^{1/2})$, the convolution
product $f*g$ is given by the equality
\[
(f*g)(\gamma)=\intop_{\gamma_{1}\circ\gamma_{2}=\gamma}f(\gamma_{1})g(\gamma_{2})
\]
with the group operation $\gamma_{1}\circ\gamma_{2}$ in $\pi$. Then
we define via $f^{*}(\gamma)=\overline{f(\gamma^{-1})}$ a $*$operation
making $C_{c}^{\infty}(\mathcal{G},\Omega^{1/2})$ into a $*$algebra.
Each level set $C_{i}$ consists of simple pieces (in case of Alexanders
horned sphere, we will explain it below) denoted by $T$. For these
pieces, one has a natural representation of $C_{c}^{\infty}(\mathcal{G},\Omega^{1/2})$
on the $L^{2}$ space over $T$. Then one defines the representation
\[
(\pi_{x}(f)\xi)(\gamma)=\intop_{\gamma_{1}\circ\gamma_{2}=\gamma}f(\gamma_{1})\xi(\gamma_{2})\qquad\forall\xi\in L^{2}(T),\forall x\in\gamma.
\]
The completion of $C_{c}^{\infty}(\mathcal{G},\Omega^{1/2})$ with
respect to the norm 
\[
||f||=\sup_{x\in\mathcal{G}}||\pi_{x}(f)||
\]
makes it into a $C^{*}$-algebra $C_{c}^{\infty}(\mathcal{G},\pi$).
Finally we are able to define the $C^{*}-$algebra associated to the
wild embedding. Using a result in \cite{TAMKrol2013}, one
can show that the corresponding von Neumann algebra is the factor
$I\! I\! I_{1}$.

Among all elements of the $C^{*}$-algebra, there are distinguished
elements, idempotent operators or projectors having a geometric interpretation.
For later use, we will construct them explicitly (we follow \cite{Connes94}
sec. $II.8.\beta$ closely). Let $Y_{\mathcal{T}}\subset\mathbb{R}^{4}$
be the wild submanifold. A small tubular neighborhood $N'$ of $Y_{\mathcal{T}}$
in $\mathbb{R}^{4}$ defines the corresponding $C^{*}$-algebra $C_{c}^{\infty}(N',\pi_{1}(\mathbb{R}^{4}\setminus N'))$
is isomorphic to $C_{c}^{\infty}(\mathcal{G},\pi_{1}(\mathbb{R}^{4}\setminus I(S^{3}))\otimes\mathcal{K}$
with $\mathcal{K}$ the $C^{*}$-algebra of compact operators. In
particular it contains an idempotent $e=e^{2}=e^{*}$, $e=1_{N}\otimes f\in C_{c}^{\infty}(\mathcal{G},\pi_{1}(\mathbb{R}^{4}\setminus I(S^{3})))\otimes\mathcal{K}$
, where $f$ is a minimal projection in $\mathcal{K}$. It induces
an idempotent in $C_{c}^{\infty}(\mathcal{G},\pi_{1}(\mathbb{R}^{4}\setminus I(S^{3})))$.
By definition, this idempotent is given by a closed curve in the complement
$\mathbb{R}^{4}\setminus I(S^{3})$. These projection operators form the basis in this algebra

\subsection{Reconstructing the Classical State}

In this section we will describe a way from a (classical) Poisson
algebra to a quantum algebra by using deformation quantization. Therefore
we will obtain a positive answer to the question: Does the $C^{*}-$algebra
of the foliation (as well of a wild (specific) embedding) comes from
a (deformation) quantization? Of course, this question cannot be answered
in all generality, but for our example we will show that the enveloping
von Neumann algebra of foliation and of this wild embedding is the
result of a deformation quantization using the classical Poisson algebra
(of closed curves) of the tame embedding. This result shows two things:
the wild embedding can be seen as a quantum state and the classical
state is a tame embedding.

\subsubsection{Intermezzo 1: The Observable Algebra and its Poisson Structure\label{sub:The-observable-algebra}}

In this section we will describe the formal structure of a classical
theory coming from the algebra of observables using the concept of
a Poisson algebra. In quantum theory, an observable is represented
by an hermitean operator having the spectral decomposition via projectors
or idempotent operators. The coefficient of the projector is the eigenvalue
of the observable or one possible result of a measurement. At least
one of these projectors represents (via the GNS representation) a quasi-classical
state. Thus, to construct the substitute of a classical observable
algebra with Poisson algebra structure, we have to concentrate on the
idempotents in the $C^{*}$-algebra. Now we will see that the set
of closed curves on a surface has the structure of a Poisson algebra.
Let us start with the definition of a Poisson algebra. 

Let $P$ be a commutative algebra with unit over $\mathbb{R}$ or
$\mathbb{C}$. A \emph{Poisson bracket} on $P$ is a bilinearform
$\left\{ \:,\:\right\} :P\otimes P\to P$ fulfilling the following
3 conditions:
\begin{enumerate}
\item anti-symmetry $\left\{ a,b\right\} =-\left\{ b,a\right\} $
\item Jacobi identity $\left\{ a,\left\{ b,c\right\} \right\} +\left\{ c,\left\{ a,b\right\} \right\} +\left\{ b,\left\{ c,a\right\} \right\} =0$
\item derivation $\left\{ ab,c\right\} =a\left\{ b,c\right\} +b\left\{ a,c\right\} $.
\end{enumerate}

Then a \emph{Poisson algebra} is the algebra $(P,\{\,,\,\})$.

Now we consider a surface $S$ together with a closed curve $\gamma$.
Additionally we have a Lie group $G$ given by the isometry group.
The closed curve is one element of the fundamental group $\pi_{1}(S)$.
From the theory of surfaces we know that $\pi_{1}(S)$ is a free abelian
group. Denote by $Z$ the free $\mathbb{K}$-module ($\mathbb{K}$
a ring with unit) with the basis $\pi_{1}(S)$, i.e. $Z$ is a freely
generated $\mathbb{K}$-module. Recall Goldman's definition of the
Lie bracket in $Z$ (see \cite{Goldman1984}). For a loop $\gamma:S^{1}\to S$
we denote its class in $\pi_{1}(S)$ by $\left\langle \gamma\right\rangle $.
Let $\alpha,\beta$ be two loops on $S$ lying in general position.
Denote the (finite) set $\alpha(S^{1})\cap\beta(S^{1})$ by $\alpha\#\beta$.
For $q\in\alpha\#\beta$ denote by $\epsilon(q;\alpha,\beta)=\pm1$
the intersection index of $\alpha$ and $\beta$ in $q$. Denote by
$\alpha_{q}\beta_{q}$ the product of the loops $\alpha,\beta$ based
in $q$. Up to homotopy the loop $(\alpha_{q}\beta_{q})(S^{1})$ is
obtained from $\alpha(S^{1})\cup\beta(S^{1})$ by the orientation
preserving smoothing of the crossing in the point $q$. Set 
\begin{equation}
[\left\langle \alpha\right\rangle ,\left\langle \beta\right\rangle ]=\sum_{q\in\alpha\#\beta}\epsilon(q;\alpha,\beta)(\alpha_{q}\beta_{q})\quad.\label{eq:Lie-bracket-loops}
\end{equation}
According to Goldman \cite{Goldman1984} (theorem 5.2), the bilinear
pairing $[\,,\,]:Z\times Z\to Z$ given by (\ref{eq:Lie-bracket-loops})
on the generators is well defined and makes $Z$ a Lie algebra.
The algebra $Sym(Z)$ of symmetric tensors is then a Poisson algebra
(see Turaev \cite{Turaev1991}).

The whole approach seems natural for the construction of the Lie algebra
$Z$ but the introduction of the Poisson structure is an artificial
act. From the physical point of view, the Poisson structure is not
the essential part of classical mechanics. More important is the algebra
of observables, i.e. functions over the configuration space forming
the Poisson algebra. For the foliation discussed above, we already
identified the observable algebra (the holonomy along closed curves)
as well the corresponding group to be $SL(2,\mathbb{C})$. Therefore
for the following, we will set $G=SL(2,\mathbb{C})$.

Now we introduce a principal $G$ bundle on $S$, representing a geometry
on the surface. This bundle is induced from a $G$ bundle over $S\times[0,1]$
having always a flat connection. Alternatively one can consider a
homomorphism $\pi_{1}(S)\to G$ represented as holonomy functional
\begin{equation}
hol(\omega,\gamma)=\mathcal{P}\exp\left(\int\limits _{\gamma}\omega\right)\in G\label{eq:holonomy-definition}
\end{equation}
with the path ordering operator $\mathcal{P}$ and $\omega$ as flat
connection (i.e. inducing a flat curvature $\Omega=d\omega+\omega\wedge\omega=0$).
This functional is unique up to conjugation induced by a gauge transformation
of the connection. Thus we have to consider the conjugation classes
of maps
\[
hol:\pi_{1}(S)\to G
\]
forming the space $X(S,G)$ of gauge-invariant flat connections of
principal $G$ bundles over $S$. Now (see \cite{Skovborg2006}) we
can start with the construction of the Poisson structure on $X(S,G)$, 
based on the Cartan form as the unique bilinearform
of a Lie algebra. As discussed above we will use the Lie group $G=SL(2,\mathbb{C})$
but the whole procedure works for every other group too. Now we consider
the standard basis
\begin{equation}
X=\left(\begin{array}{cc}
0 & 1\\
0 & 0
\end{array}\right)\quad,\qquad H=\left(\begin{array}{cc}
1 & 0\\
0 & -1
\end{array}\right)\quad,\qquad Y=\left(\begin{array}{cc}
0 & 0\\
1 & 0
\end{array}\right)\label{eq:generators-SL2C}
\end{equation}
of the Lie algebra $sl(2,\mathbb{C})$ with $[X,Y]=H,\,[H,X]=2X,\,[H,Y]=-2Y$.
Furthermore there is the bilinearform $B:sl_{2}\otimes sl_{2}\to\mathbb{C}$
written in the standard basis as 
\[
\left(\begin{array}{ccc}
0 & 0 & -1\\
0 & -2 & 0\\
-1 & 0 & 0
\end{array}\right)
\]
Now we consider the holomorphic function $f:SL(2,\mathbb{C})\to\mathbb{C}$
and define the gradient $\delta_{f}(A)$ along $f$ at the point $A$
as $\delta_{f}(A)=Z$ with $B(Z,W)=df_{A}(W)$ and 
\[
df_{A}(W)=\left.\frac{d}{dt}f(A\cdot\exp(tW))\right|_{t=0}\quad.
\]
The calculation of the gradient $\delta_{tr}$ for the trace $tr$
along a matrix 
\[
A=\left(\begin{array}{cc}
a_{11} & a_{12}\\
a_{21} & a_{22}
\end{array}\right)
\]
 is given by
\[
\delta_{tr}(A)=-a_{21}Y-a_{12}X-\frac{1}{2}(a_{11}-a_{22})H\quad.
\]
Given a representation $\rho\in X(S,SL(2,\mathbb{C}))$ of the fundamental
group and an invariant function $f:SL(2,\mathbb{C})\to\mathbb{R}$
extendable to $X(S,SL(2,\mathbb{C}))$. Then we consider two conjugacy
classes $\gamma,\eta\in\pi_{1}(S)$ represented by two transversal
intersecting loops $P,Q$ and define the function $f_{\gamma}:X(S,SL(2,\mathbb{C})\to\mathbb{C}$
by $f_{\gamma}(\rho)=f(\rho(\gamma))$. Let $x\in P\cap Q$ be the
intersection point of the loops $P,Q$ and $c_{x}$ a path between
the point $x$ and the fixed base point in $\pi_{1}(S)$. Then we
define $\gamma_{x}=c_{x}\gamma c_{x}^{-1}$ and $\eta_{x}=c_{x}\eta c_{x}^{-1}$.
Finally we get the Poisson bracket
\[
\left\{ f_{\gamma},f'_{\eta}\right\} =\sum_{x\in P\cap Q}sign(x)\: B(\delta_{f}(\rho(\gamma_{x})),\delta_{f'}(\rho(\eta_{x})))\quad,
\]
where $sign(x)$ is the sign of the intersection point $x$. Thus,

\emph{The space $X(S,SL(2,\mathbb{C}))$ has a natural Poisson structure
(induced by the bilinear form (\ref{eq:Lie-bracket-loops}) on the
group) and the Poisson algebra }$(X(S,SL(2,\mathbb{C}),\left\{ \,,\,\right\} )$\emph{
of complex functions over them is the algebra of observables.}

\subsubsection{Intermezzo 2: Drinfeld-Turaev Quantization\label{sub:Drinfeld-Turaev-Quantization}}

Now we introduce the ring $\mathbb{C}[[h]]$ of formal polynomials
in $h$ with values in $\mathbb{C}$. This ring has a topological
structure, i.e. for a given power series $a\in\mathbb{C}[[h]]$ the
set $a+h^{n}\mathbb{C}[[h]]$ forms a neighborhood. Now we define 

A \emph{Quantization} of a Poisson algebra $P$ is a $\mathbb{C}[[h]]$
algebra $P_{h}$ together with the $\mathbb{C}$-algebra isomorphism
$\Theta:P_{h}/hP\to P$ so that

1. the module $P_{h}$ is isomorphic to $V[[h]]$ for a $\mathbb{C}$
vector space $V$

2. let $a,b\in P$ and $a',b'\in P_{h}$ be $\Theta(a)=a'$, $\Theta(b)=b'$
then
\[
\Theta\left(\frac{a'b'-b'a'}{h}\right)=\left\{ a,b\right\} \;.
\]

One speaks of a deformation of the Poisson algebra by using a deformation
parameter $h$ to get a relation between the Poisson bracket and the
commutator. Therefore we have the problem to find the deformation
of the Poisson algebra $(X(S,SL(2,\mathbb{C})),\left\{ \,,\,\right\} )$.
The solution to this problem can be found via two steps: 
\begin{enumerate}
\item at first find another description of the Poisson algebra by a structure
with one parameter at a special value and 
\item secondly vary this parameter to get the deformation. 
\end{enumerate}
Fortunately both problems were already solved (see \cite{Turaev1989,Turaev1991}).
The solution of the first problem is expressed in the theorem: 

\emph{The skein module $K_{-1}(S\times[0,1])$ (i.e. $t=-1$) has
the structure of an algebra isomorphic to the Poisson algebra $(X(S,SL(2,\mathbb{C})),\left\{ \,,\,\right\} )$.}
\emph{(see also \cite{Bullock1999,BulPrzy:1999}) }

Then we have also the solution of the second problem: 

\emph{The skein algebra $K_{t}(S\times[0,1])$ is the quantization
of the Poisson algebra $(X(S,SL(2,\mathbb{C})),\left\{ \,,\,\right\} )$
with the deformation parameter $t=\exp(h/4)$.(see also \cite{BulPrzy:1999})} 

\begin{figure}
\begin{center}\includegraphics[scale=0.25]{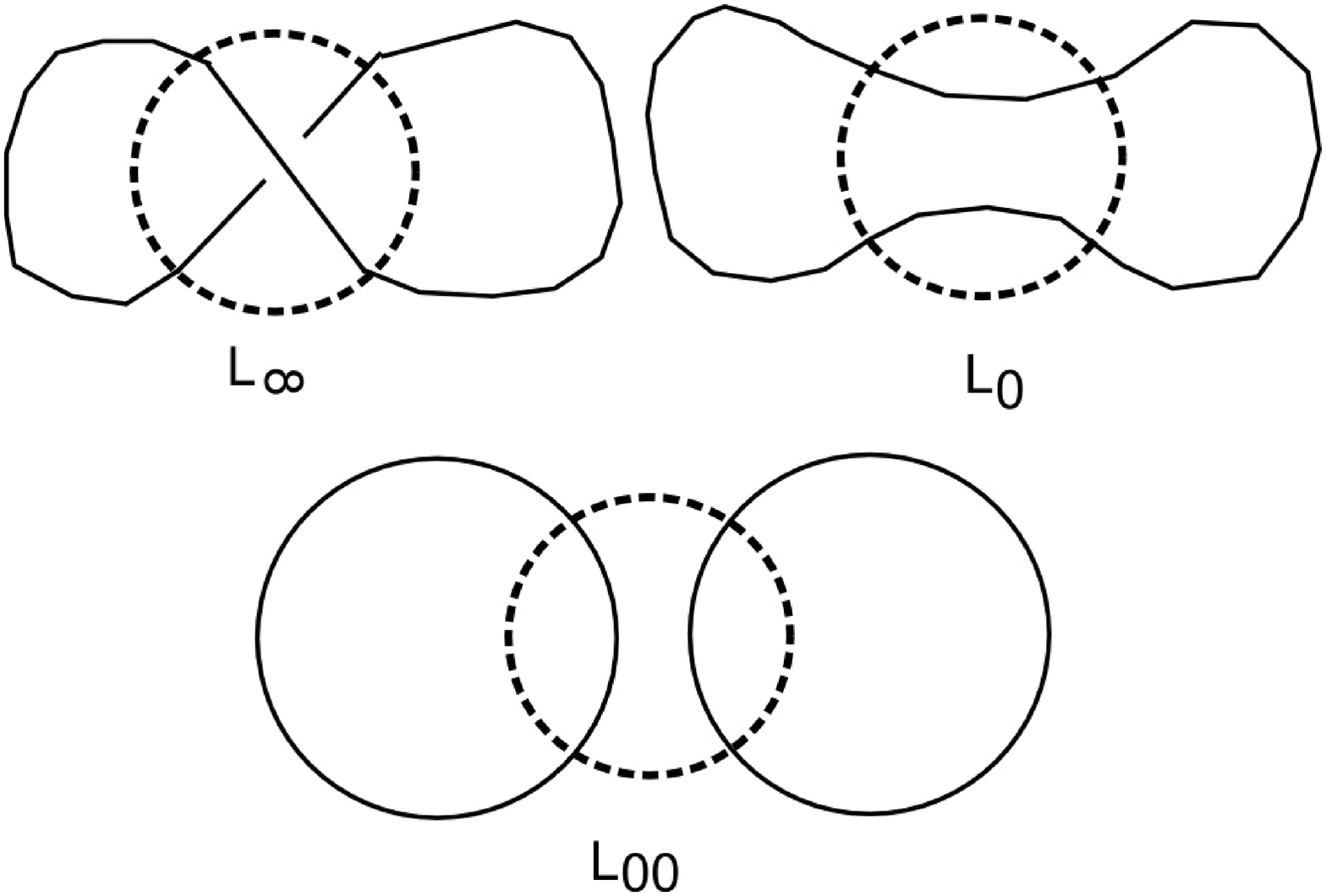}\end{center}
\caption{Crossings $L_{\infty},L_{o},L_{oo}$.\label{fig:skein-crossings}}
\end{figure}

To understand these solutions we have to introduce the skein module
$K_{t}(M)$ of a 3-manifold $M$ (see \cite{PrasSoss:97}). For that
purpose we consider the set of links $\mathcal{L}(M)$ in $M$ up
to isotopy and construct the vector space $\mathbb{C}\mathcal{L}(M)$
with basis $\mathcal{L}(M)$. Then one can define $\mathbb{C}\mathcal{L}[[t]]$
as ring of formal polynomials having coefficients in $\mathbb{C}\mathcal{L}(M)$.
Now we consider the link diagram of a link, i.e. the projection of
the link to the $\mathbb{R}^{2}$ having the crossings in mind. Choosing
a disk in $\mathbb{R}^{2}$ so that one crossing is inside this disk.
If the three links differ by the three crossings $L_{oo},L_{o},L_{\infty}$
(see figure \ref{fig:skein-crossings}) inside of the disk then these
links are skein-related. 

Then in $\mathbb{C}\mathcal{L}[[t]]$ one writes the skein relation%
\footnote{The relation depends on the group $SL(2,\mathbb{C})$.%
} $L_{\infty}-tL_{o}-t^{-1}L_{oo}$. Furthermore let $L\sqcup O$ be
the disjoint union of the link with a circle and one writes the framing
relation $L\sqcup O+(t^{2}+t^{-2})L$. Let $S(M)$ be the smallest
submodule of $\mathbb{C}\mathcal{L}[[t]]$ containing both relations.
Then we define the Kauffman bracket skein module by $K_{t}(M)=\mathbb{C}\mathcal{L}[[t]]/S(M)$.
We list the following general results about this module:
\begin{itemize}
\item The module $K_{-1}(M)$ for $t=-1$ is a commutative algebra.
\item Let $S$ be a surface, then $K_{t}(S\times[0,1])$ carries the structure of an algebra.
\end{itemize}
The algebraic structure of $K_{t}(S\times[0,1])$ can be simply seen
by using the diffeomorphism between the sum $S\times[0,1]\cup_{S}S\times[0,1]$
along $S$ and $S\times[0,1]$. Then the product $ab$ of two elements
$a,b\in K_{t}(S\times[0,1])$ is a link in $S\times[0,1]\cup_{S}S\times[0,1]$
corresponding to a link in $S\times[0,1]$ via the diffeomorphism.
The algebra $K_{t}(S\times[0,1])$ is in general non-commutative for
$t\not=-1$. For the following we will omit the interval $[0,1]$
and denote the skein algebra by $K_{t}(S)$.

In subsection \ref{sub:Construction-of-state}, we described the state
as an element of the Kauffman skein module $K_{t}(\ell)$ of the leaf
$\ell$. Now we obtained also that the observable algebra is the Kauffman
skein module again. How does this whole story fit into the description
of the observable algebra for the foliation as factor $I\! I\! I_{1}$?
In \cite{FrohmanGelca2000}, it was shown that the Kauffman bracket skein module of
a cylinder over the torus embeds as a subalgebra of the noncommutative
torus. However, the noncommutative torus can be seen as the leaf space
of the Kronecker foliation of the torus leading to the factor $I\! I_{\infty}$.
Then by using (\ref{eq:crossed-product-factor-III}), we obtain the
factor $I\! I\! I_{1}$ back. We will use this relation in the next
section to get the quantum action.

\section{Action at the Quantum Level\label{sec:Action-at-the-quantum-level}}

Above, we used the foliation to get quantum states which agreed with
the deformation quantization of a classical state. Central point in
our argumentation is the construction of the $C^{*}-$algebra with
the corresponding von Neumann algebra as observable algebra. This
von Neumann algebra is a factor $I\! I\! I_{1}$. By using the Tomita-Takesaki
modular theory, there is a relation to the factor $I\! I_{\infty}$
by using an action of the group $\mathbb{R}_{+}^{*}$ by automorphisms
of a Lebesgue measure space leading to the decomposition of the factor
$I\! I\! I_{1}$. This action is related to an invariant, the flow
of weights mod(M). The main property of the factor $I\! I\! I_{1}$
is the constant flow of weights mod(M). Connes \cite{Connes1984,Connes94}
described the flow of weights as a bundle of densities over the leaf
space, i.e. the $\mathbb{R}_{+}^{*}$ homogeneous space of nonzero
maps. In case of foliation considered above, this density is constant
and we can naturally identify this density with the volume of the
submanifold defining the foliation. By definition, this volume
is given by the Godbillon-Vey invariant (see eqn. (\ref{eq:GV-number-Thurston-foliation-1})
in Appendix B, the circle in the fiber has unit size). This invariant
can be seen as an element of $H^{3}(BG,\mathbb{R})$ with the holonomy
groupoid $G$ of the foliation. As shown by Connes \cite{Connes1984,Connes94},
the Godbillon-Vey class $GV$ can be expressed as a cyclic cohomology
class (the so-called flow of weights)
\[
GV_{HC}\in HC^{2}(C_{c}^{\infty}(G))
\]
of the $C^{*}-$algebra for the foliation. Then we define an expression
\[
S=Tr_{\omega}\left(GV_{HC}\right)
\]
uniquely associated to the foliation ($Tr_{\omega}$ is the Dixmier
trace). The expression $S$ generates the action on the factor by
\[
\Delta_{\omega}^{it}=exp(i\, S)
\]
so that $S$ is the action or the Hamiltonian multiplied by the time
(see (\ref{eq:modular-generator})). It is an operator which defines
the dynamics by acting on the states. For explicit calculations we
have to evaluate this operator. One way is the usage of the relation
between the foliation and the wild embedding. This wild embedding
is determined by the fundamental group $\pi$ of its complement. In
\cite{TAMKrol2013}, we discussed the properties of this group
$\pi$. It is a perfect group, i.e. every element is generated by
a commutator. Then a representation of this group into some other
group like $GL(\mathbb{C})$ (the limit of $GL(n,\mathbb{C})$ for
$n\to\infty$) reduces to the representation of the maximal perfect
subgroup. For that purpose we consider the representation of the group
$\pi$ into the group $E(\mathbb{C})$ of elementary matrices, which
is the perfect subgroup of $GL(\mathbb{C})$. Then we obtain matrix-valued
functions $X^{\mu}\in C_{c}^{\infty}(E(\mathbb{C}))$ as the image
of the generators of $\pi$ w.r.t. the representation $\pi\to E(\mathbb{C})$
labeled by the dimension $\mu=1,\ldots,4$ of the embedding space
$R^{n}$. Via the representation $\iota:\pi\to E(\mathbb{C})$, we
obtain a cyclic cocycle in $HC^{2}(C_{c}^{\infty}(E(\mathbb{C}))$
generated by a suitable Fredholm operator $F$. Here we use the standard
choice $F=D|D|^{-1}$ with the Dirac operator $D$ acting on the functions
in $C_{c}^{\infty}(E(\mathbb{C}))$. Then the cocycle in $HC^{2}(C_{c}^{\infty}(E(\mathbb{C}))$
can be expressed by
\[
\iota_{*}GV_{HC}=\eta_{\mu\nu}[F,X^{\mu}][F,X^{\nu}]
\]
using a metric $\eta_{\mu\nu}$in $R^{4}$ via the pull-back using
the representation $\iota:\pi\to E(\mathbb{C})$. Finally we obtain
the action
\begin{equation}
S=Tr_{\omega}([F,X^{\mu}][F,X_{\mu}])=Tr_{\omega}([D,X^{\mu}][D,X_{\mu}]|D|^{-2})\label{eq:quantum-action}
\end{equation}
which can be evaluated by using the heat-kernel of the Dirac operator
$D$. The appearance of the heat kernel is a sign for a relation to
quantum field theory where the heat kernel is a very convenient tool
for studying one-loop divergences, anomalies and various asymptotics
of the effective action. 

Away from this operator expression for the Godbillon-Vey invariant,
there are geometrical evaluations which are not defined on the leaf
space but rather on the whole manifold. As mentioned above, this invariant
admits values in the real numbers and we can evaluate them according
to the type of the number: for integer values one obtains the Euler
class and for rational numbers the Pontrjagin class (for the corresponding
bundles). Therefore using the ideas of section \ref{sec:action-induced-by-top},
we obtain the Einstein-Hilbert and the Holst action but also a correction
given by irrational values of the Godbillon-Vey number.

\section{The Scaling Behavior of the Action}

A good test for the theory is the dependence of the action (\ref{eq:quantum-action})
on the scale. The theory has a strong geometrical flavor and therefore
the scaling behavior can be understood by a geometrical construction
using the exotic $R^{4}$. As explained above, the central point in
the construction is the Casson handle. From the scaling point of view,
the Casson handle contains disks of any size (with respect to the
embedding $R^{4}\hookrightarrow\mathbf{R^{4}}$). The long scales
are given by the first levels of the Casson handle whereas the small
scales are represented by the higher levels of the Casson handle.

\subsection{Long-scale Behavior (Einstein-Hilbert Action)\label{sub:Long-scale-behaviour}}

Let us consider the small exotic $R^{4}$. From the physics point
of view, the large scale is given by the first levels of the Casson
handle. In the construction of the foliation of $R^{4}$, the first
levels describe a polygon in the hyperbolic space $\mathbb{H}^{2}$
with a finite and small number of vertices. The Godbillon-Vey number
of this foliation is given by the volume of this polygon. In principle,
it is also true for the inclusion of the higher levels (and also for
the whole Casson handle) but every higher level gives only a very
small contribution to the Godbillon-Vey number. Therefore, the first
levels of the Casson handle can be simply characterized by the Godbillon-Vey
number, i.e. by the size of the polygon in the scale $r$. Then the
Godbillon-Vey number is given by $GV=r^{2}$. In \cite{TAMKrol2013}
we analyzed this situation and found the relation 
\[
GV\stackrel{r\to\infty}{=}  r^{2}\intop_{D^{2}}\left(g_{\mu\nu}\partial_{k}\xi^{\mu}\partial^{k}\xi^{\nu}\right)d^{2}x
\]
to the Godbillon-Vey number. Here we integrate over the disk (equal
to the polygon) which is used to define the foliation. This model
is the non-linear sigma model (for the embedding of the disk into
$Y_{n}$ with metric $g$) depending on the scale $r^{2}$. The scaling
behavior of this model was studied in \cite{Friedan1980} and one
obtains the RG flow equation 
\begin{equation}
\frac{\partial}{\partial r^{2}}g_{\mu\nu}=R_{\mu\nu}+\frac{1}{r^{2}}\left(R_{\mu\lambda\kappa\iota}R_{\nu}^{\lambda\kappa\iota}\right)+O(r^{-4})\label{eq:RG-flow-equation}
\end{equation}
reducing to the Ricci flow equations for large scales ($r\to\infty$).
The fixed point of this flow are geometries of constant curvature (used
to prove the Thurston geometrization conjecture). Therefore in the
classical limit of large scales, we obtain a geometry of the 3-manifold
of constant curvature whereas for small scales one has to take into
account higher curvature corrections. On the spacetime, one has also
flow equations from one 3-manifold of constant curvature to another
3-manifold of constant curvature. This flow equation is equivalent
to the (anti-)self-dual curvature (or instantons) by using the gradient
flow of the Chern-Simons functional \cite{Flo:88}. This approach
has much in common with the non-linear graviton of Penrose \cite{Penrose1976}.
We will explain these ideas in subsection \ref{sub:The-Graviton-propagator}.

\subsection{Short-scale Behavior\label{sub:Short-scale-behaviour}}

For the short scale, we need the full power of the Casson handle.
As a first step we can evaluate the action (\ref{eq:quantum-action})
so that the Dirac operator $D$ acts on usual square-integrable functions,
so that $[D,X^{\mu}]=dX^{\mu}$ is finite. Then the action (\ref{eq:quantum-action})
reduces to 
\[
S=Tr_{\omega}(\eta_{\mu\nu}(\partial_{k}X^{\mu}\partial^{k}X^{\nu})|D|^{-2})
\]
where $\mu,\nu=1,\ldots,4$ is the index for the coordinates on $R^{4}$
and $k=1,2$ represents the index of the disk (inside of the Casson
handle). Now we will choose a small fluctuation $\xi^{k}$ of a fixed
embedding of the disk in the Casson handle given by $X^{\mu}=(x^{k}+\xi^{\mu})\delta_{k}^{\mu}$
with $\partial_{l}x^{k}=\delta_{l}^{k}$. Then we obtain
\[
\partial_{k}X^{\mu}\partial^{k}X^{\nu}=\delta_{k}^{\mu}\delta_{k}^{\nu}(1+\partial_{k}\xi^{\mu})(1+\partial_{k}\xi^{\nu})
\]
and we use a standard argument to neglect the terms linear in $\partial\xi$:
fluctuations have no preferred direction and therefore only the square
contributes. Then we have 
\[
S=Tr_{\omega}(\eta_{\mu\nu}(\delta_{k}^{\mu}\delta_{k}^{\nu}+\partial_{k}\xi^{\mu}\partial^{k}\xi^{\nu})|D|^{-2})
\]
for the action. By using a result of \cite{Connes94} one obtains
for the Dixmier trace
\[
Tr_{\omega}(|D|^{-2})=2\intop_{D^{2}}*(\Phi_{1})
\]
with the first coefficient $\Phi_{1}$ of the heat kernel expansion
\cite{bvg:90}
\[
\Phi_{1}=\frac{1}{6}R
\]
and the action simplifies to
\begin{equation}
S=\intop_{D^{2}}\left(\frac{2}{3}R+\partial_{k}\xi^{\mu}\partial^{k}\xi^{\nu}\frac{1}{3}R\right)dvol(D^{2})\label{eq:action-small-fluctuations}
\end{equation}
for the main contributions where $R$ is the scalar curvature of the
embedded disk $D^{2}$. Again, but now for small fluctuations, we obtain
the flow equation (\ref{eq:RG-flow-equation}) but we have to consider
the small case $r\to0$. Then we have to take arbitrary curvature
contributions into account. This short calculation showed that the
short-scale behavior is given by a two-dimensional action. In the
next section we will understand this behavior geometrically. For small
fluctuations we obtained a disk but what happens for larger fluctuations?
Then we have to take even the higher levels of the Casson handle into
account. These higher levels form a complicated surface with a
fractal structure (a generalization of the Cantor set). Then the action
(\ref{eq:action-small-fluctuations}) has to be replaced by an integral
over this fractal space. For the evaluation of the quantum action
(\ref{eq:quantum-action}) one can use the ideas of noncommutative
geometry as used for fractals and quasi-Fuchsian groups, see \cite{Connes94} (section IV.3).

\section{Some Properties of the Theory\label{sec:Some-Properties}}

In this section we will present some properties of the theory. For
an impression, it is enough to present the main ideas. The details will be published separately.

\subsection{The Graviton\label{sub:The-Graviton-propagator}}

By using the large scale behavior in subsection \ref{sub:Long-scale-behaviour},
we have to consider Ricci-flat spaces and an easy calculation gives
the well-known propagator in the linearized version, however, we are not
interested in the linearized version. GRT is a highly non-linear theory
and therefore one has to take this non-linearity into account. The
Ricci-flatness of the spacetime goes over to the 3-manifold as the
spatial component where it implies a 3-manifold of constant curvature
(as fixed point of the Ricci flow). Then as shown by Witten \cite{Wit:89.2,Wit:89.3,Wit:91.2},
the 3-dimensional Einstein-Hilbert action
\[
\intop_{N}R_{(3)}\sqrt{h}\, d^{3}x=L\cdot CS(N,A)
\]
is related to the Chern-Simons action $CS(N,A)$ with respect to the
(Levi-Civita) connection $A$ and the length $L$. By using the Stokes
theorem we obtain
\[
S_{EH}(N\times[0,1])=\intop_{M_{T}}tr(F\wedge F) \, ,
\]
i.e. the action for the 4-manifold $N\times[0,1]$ (as local spacetime)
with the curvature $F=DA$, i.e. the action is the (topological) Pontrjagin
class of the 4-manifold. From the formal point of view, the curvature
2-form $F=DA$ is generated by a $SO(3,1)$ connection $A$ in the
frame bundle, which can be lifted uniquely to a $SL(2,\mathbb{C})$-
(Spin-) connection. According to the Ambrose-Singer theorem, the components
of the curvature tensor are determined by the values of holonomy which
is in general a subgroup of $SL(2,\mathbb{C})$. Thus we start with
a suitable curvature 2-form $F=DA$ with values in the Lie algebra
$\mathfrak{g}$ of the Lie group $G$ as subgroup of the $SL(2,\mathbb{C})$.
The variation of the Chern-Simons action gets flat connections $DA=0$
as solutions. The flow of solutions $A(t)$ in $N\times[0,1]$ (parametrized
by the variable $t$, the 'time') between the flat connection $A(0)$
in $N\times\left\{ 0\right\} $ to the flat connection $A(1)$ in
$N\times\left\{ 1\right\} $ will be given by the gradient flow equation
(see for instance \cite{Flo:88})
\begin{equation}
\frac{d}{dt}A(t)=\pm*F(A)=\pm*DA\label{eq:gradient-flow}
\end{equation}
where the coordinate $t$ is normal to $N$. Therefore we are able
to introduce a connection $\tilde{A}$ in $N\times[0,1]$ so that
the covariant derivative in $t$-direction agrees with $\partial/\partial t$.
Then we have for the curvature $\tilde{F}=D\tilde{A}$, where the fourth
component is given by $\tilde{F}_{4\mu}=d\tilde{A}_{\mu}/dt$. Thus
we will get the instanton equation with (anti-) self-dual curvature
\[
\tilde{F}=\pm*\tilde{F}\,.
\]
It follows
\[
S_{EH}([0,1]\times N)=\intop_{N\times[0,1]}tr(\tilde{F}\wedge\tilde{F})=\pm\intop_{N\times[0,1]}tr(\ \tilde{F}\wedge*\tilde{F})\,,
\]
(i.e. the MacDowell\textendash{}Mansouri action). 

We remark the main point in this argumentation: we obtain a self-dual curvature as gradient flow between two 3-manifolds of constant curvature. Of course, (anti-)self-dual curvatures are also solutions of Einsteins equation (but the reverse is not true). Following Penrose \cite{Penrose1976}, we call these solutions the nonlinear graviton.

\subsection{Relations to the Quantum Groups\label{sub:Relations-to-quantum-group}}

Above we constructed the observable algebra from the foliation leading
to the Kauffman bracket skein module. In the subsection we will discuss
the relation to lattice gauge field theory. Main source
for this discussion is the work of Bullock, Frohman and Kania-Bartoszy{\'n}ska
\cite{BullockFrohmanKania-Bartoszynska1998,BullockFrohmanKania-Bartoszynska1999,BullockFrohmanKania-Bartoszynska2002,FrohmanGelca2000}.
In this paper the authors realize that gauge fields come from the
restricted dual of the Hopf algebra on which the theory is based.
This leads to a coordinate free formulation. Then they comultiply
connections in a way that implies the usual exchange relations for
fields while preserving their evaluability. Their new foundations
allow them to compute Wilson loops and many other operators using
a simple extension of tangle functors. Then they analyzed the structure
of the algebra of observables. In their viewpoint, the observables
correspond to quantum groups seen as rings of invariants of n-tuples
of matrices under conjugation. The connection with lattice gauge field
theory is that each n-tuple of matrices corresponds to a connection
on a lattice with one vertex and n-edges, with the gauge fields based
on a classical group. The construction given in this paper leads to
an algebra of ''characters'' of a surface group with respect to
any ribbon Hopf algebra. The algebras are interesting from many points
of view: They generalize objects studied in invariant theory; they
should provide tools for investigating the structure of the mapping
class groups of surfaces; and they should give a way of understanding
quantum invariants of $3$-manifolds. The algebra of observables based
on the enveloped Lie algebra $U(\mathfrak{g})$ is proved to be the
ring of $G$-characters of the fundamental group of the associated
surface. Then, given the ring of $G$-characters of a surface group,
they showed that the observables based on the corresponding Drinfeld-Jimbo
algebra form a quantization with respect to the usual Poisson structure.
Furthermore they proved for the classical groups that the algebra
of observables is generated by Wilson loops. Finally, invoking a quantized
Cayley-Hamilton identity, they obtain a new proof, that the $U_{h}(sl_{2})$-characters
of a surface are exactly the Kauffman bracket skein module of a cylinder
over that surface. The power of lattice gauge field theory is that
it places the representation theory of the underlying manifold and
the quantum invariants in the same setting. Ultimately the asymptotic
analysis of the quantum invariants of a $3$-manifold in terms of
the representations of its fundamental group should flow out of this
setting. The identification of the representation theory of a quantum
group with that of a compact Lie group leads to rigorous integral
formulas for quantum invariants of $3$-manifolds. This should in
turn lead to a simple explication of the relationship between quantum
invariants and more classical invariants of $3$-manifolds.

This relation to lattice gauge field theory seems to imply an underlying
discrete structure of the space and/or spacetime, but the approach in the 
paper uncovers the reason \cite{BullockFrohmanKania-Bartoszynska1998,BullockFrohmanKania-Bartoszynska1999,BullockFrohmanKania-Bartoszynska2002,FrohmanGelca2000}:
the Kauffman bracket skein module is discrete structure containing
only a finite amount of information. Therefore, any description has to be discrete as well
including the approach via gauge fields. This idea can be extended
to the 4-manifold. As explained above, every smooth 4-manifold can
be effectively described by handles and one only needs a finite number
to describe every compact 4-manifold. Then the handles can be simply
triangulated by using simplices to end up with a piecewise-linear
(or PL) structure. The surprising result of Cerf for manifolds of
dimension smaller than 7 was simple: PL-structure (or triangulations)
and smoothness structure are the same. This implies that every PL-structure
can be smoothed to a smoothness structure and vice versa. Therefore
\emph{the discrete approach (via triangulations) and the smooth approach
to defining a manifold are the same}! So, our spacetime admits a kind
of duality: it contains discrete information in its handle structure
but it is a continuous space at the same time. Both approaches are
interchangeable. Therefore the underlying structure of the spacetime
is discrete but the spacetime itself is a smooth 4-manifold. Or,
the information contained in a smooth 4-manifold is finite.

\subsection{Dimensional Reduction and Exotic Smooth Black Holes\label{sub:Dimensional-reduction}}

In \cite{TAMBrans2012} we describe an exotic black hole by
constructing a smooth metric for the interior. Here we will present
the main argument shortly.

In \cite{Brans:94a} the existence of an exotic Black hole (as exotic
Kruskal space) using an exotic $\mathbb{R}^{4}$ was suggested. The
idea was simply to consider the complement $\mathbb{R}^{4}\setminus(D^{3}\times\mathbb{R})=S^{2}\times\mathbb{R}^{2}$
where $\times$ was only understood topologically. In case of the
exotic small $\mathbb{R}^{4}$ given by a Casson handle, we can reproduce
our construction of an exotic $S^{2}\times\mathbb{R}^{2}$ by using
a Casson handle. Therefore we will here concentrate on the representation
of the exotic $S^{2}\times\mathbb{R}^{2}$ by using the Casson handle
$CH$ to get 
\[
S^{2}\times_{\Theta}\mathbb{R}^{2}=D^{2}\cup_{\partial CH}CH\:.
\]
 In \cite{Kato:2004} the analytical properties of the Casson handle
were discussed. The main idea is the usage of the theory of end-periodic
manifolds, i.e. an infinite periodic structure generated by $W$ glued
along a compact set $K$ to get 
\[
S^{2}\times_{\Theta}\mathbb{R}^{2}=K\cup_{N}W\cup_{N}W\cup_{N}\cdots
\]
 the end-periodic manifold. The definition of an end-periodic manifold
is very formal (see \cite{Tau:87}) and we omit it here. All Casson
handles generated by a balanced tree have the structure of end-periodic
manifolds as shown in \cite{Kato:2004}. By using the theory of Taubes
\cite{Tau:87} one can construct a metric on $\cdots\cup_{N}W\cup_{N}W\cup_{N}\cdots$
by using the metric on $W$. Then a metric $g$ in $S^{2}\times_{\Theta}\mathbb{R}^{2}$
transforms to a periodic function $\hat{g}$ on the infinite periodic
manifold
\[
\tilde{Y}=\cdots\cup_{N}W_{-1}\cup_{N}W_{0}\cup_{N}W_{1}\cup_{N}\cdots
\]
 where $W_{i}$ is the building block $W$ at the $i$-th place. To
reflect the number of the building block, we have to extend $\hat{g}$
to $Y\times\mathbb{C}^{*}$by using a metric $\hat{g}_{z}$holomorphic
in $z\in\mathbb{C}^{*}=S^{1}$ with $Y=W/i$ where $i$ identifies
the two boundaries of $W$. From the formal point of view we have
the \emph{generalized Fourier-Laplace transform (or Fourier-Laplace
transform for short)}
\begin{equation}
\hat{g}_{z}(.)=\sum_{n=0}^{\infty}a_{n}z^{n}\cdot\hat{g}(.)\label{eq:periodic-metric-on-YxS1}
\end{equation}
 where the coefficient $a_{n}$ represents the building block $W_{n}$
in $\tilde{Y}$. Without loss of generality we can choose the coordinates
$x$ in $M$ so that the $0$-th component $x_{0}$ is related to the
integer $n=[x_{0}]$ via its integer part $[\:]$. Using the inverse
transformation we can construct a smooth metric $g$ in $\tilde{Y}$
at the $n$-th building block via
\begin{equation}
(\tilde{T}^{n}g)(x)=\frac{1}{2\pi i}\intop_{|z|=s}z^{-n}\hat{g}_{z}(\pi(x))\frac{dz}{z}\label{eq:FL-trafo}
\end{equation}
 for $x\in\tilde{Y}$, $s\in(0,\infty)$, $n=[x_{0}]$ with the projection
$\pi:\tilde{Y}\to Y$ (mathematically: $\tilde{Y}$ is the universal
cover of $Y$ like $\mathbb{R}$ is the universal cover of $S^{1}$).
In the case of the Kruskal space we have the metric 
\begin{equation}
ds^{\text{2}}=\left(\frac{2M^{3}}{r}\right)\exp\left(-\frac{r}{2M}\right)\left(-dv^{2}+du^{2}\right)+r^{2}\left(d\theta^{2}+\sin^{2}\theta d\phi^{2}\right)\label{eq:Kruskal-metric}
\end{equation}
 in the usual units with a singularity at $r=0$ used for the whole
space $S^{2}\times\mathbb{R}^{2}$. The coordinates $(u,v)$ together
with the relation
\begin{equation}
u^{2}-v^{2}=\left(\frac{r}{2M}-1\right)\exp\left(-\frac{r}{2M}\right)\label{eq:Kruskal-relation}
\end{equation}
 represent $\mathbb{R}^{2}$ and the angles $(\theta,\phi)$ the 2-sphere
$S^{2}$ parametrized by the radius $r$. Clearly this metric can
be also used for each building block $W$ having the topological structure
$D^{3}\times S^{1}=W$ with two attaching regions topologically given
by $D^{2}\times S^{1}=N$ forming the boundary $S^{2}\times S^{1}=\partial(D^{3}\times S^{1})$
(see the description of a Casson handle above). Remember that the
Casson handle is topologically the subset $D^{2}\times\mathbb{R}^{2}\subset S^{2}\times\mathbb{R}^{2}$.
Now we consider the decomposition $W=D^{2}\times(D^{1}\times S^{1})$
and the part $D^{1}\times S^{1}$ will be later the $\mathbb{R}^{2}$
part of the Casson handle. The size of the $D^{2}$ is parametrized
by $r$ as above. Then we obtain the metric (\ref{eq:Kruskal-metric})
for the building block $W$.

\begin{figure}
\begin{center}
\includegraphics[scale=0.4]{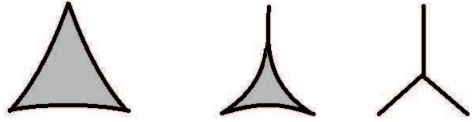}
\caption{Hyperbolic triangle with increasing curvature (from left to right), the tree is the limit  and a dimensional reduction 2D to 1D\label{fig:degen-hyp-tree}}
\end{center}
\end{figure}

Our model of the black hole based on the implicit dependence of the
two coordinates $(u,v)$ on the parameter $r$, the radius of the
2-sphere. Therefore we choose for the coordinate $z\in\mathbb{C}^{*}$
the relation $z=\exp(ir)$ and obtain a metric $\hat{g}$ on $Y\times\mathbb{C}^{*}$.
So, we make the assumptions: 
\begin{enumerate}
\item The coordinate $z$ is related to the radius by $z=\exp(ir)$. 
\item Only the $(u,v)$ part of the metric is periodic and we do not change
the other component $r^{2}\left(d\theta^{2}+\sin^{2}\theta d\phi^{2}\right)$
of the metric. 
\item The integer part $n=[v]$ of the coordinate $v$ gives the number
of the building block $W_{n}$ in the Casson handle (seen as end-periodic
manifold). 
\item The metric on $S^{2}\times_{\theta}\mathbb{R}^{2}$ is given by a
Fourier transformation (\ref{eq:FL-trafo}) of the $(u,v)$ part of
the metric in the building block $W$. 
\end{enumerate}
Some more comments are in order. The number $n=[v]$ is related to
the coordinate $v$ as substitute of ''time''. The metric $g$ in
$\tilde{Y}$ is smooth with respect to $v$ and we obtain the number
of the building block by $n=[v]$. To express this property we have
to identify $(u,v)$ with the coordinates of $D^{1}\times S^{1}$.
Then we obtain the metric on $S^{2}\times_{\theta}\mathbb{R}^{2}$
by the generalized Fourier-Laplace transformation of the metric on
$Y=W/i$ using the metric of the building block $W$ and the coordinate
$z$ similar to (\ref{eq:FL-trafo})
\[
g(v,u,\theta,\phi)=\intop\exp\left(irv\right)\hat{g}_{r}(v,u,\theta,\phi)dr
\]
 Especially the singular part of the metric (i.e. the $(u,v)$ part)
on the building block $W$
\[
(\hat{g}_{r})_{00}=\left(\frac{2M^{3}}{r}\right)\exp\left(-\frac{r}{2M}\right)=(\hat{g}_{r})_{11}
\]
 transforms to the Heaviside jump function
\[
g_{00}=g_{11}=2M^{3}\Theta(v^{2}-u^{2}-1)
\]
 using the relation (\ref{eq:Kruskal-relation}), having no singularity.
The metric vanishes, however, for large values of $v$ in the interior of
the black hole. This sketch of some arguments gives a hint that the
transformation of the smoothness to exotic smoothness could possibly
smooth out some singularity in the black hole case. This metric vanishes
along two directions or \emph{one obtains a dimensional reduction from 4D to 2D.} 
But, is there a geometrical reason for this reduction? A hyperbolic
3-manifold $M$ admits a hyperbolic structure by fixing a homomorphism
$\pi_{1}(M)\to SL(2,\mathbb{C})$ (up to conjugation). From the physics
point of view, this homomorphism is given by the holonomy along a
closed curve (as element in $\pi_{1}(M)$) for a flat connection.
A sequence of these holonomies does not converge but it is possible
to compactify the space of flat $SL(2,\mathbb{C})$ connections. This
limit can be understood geometrically: the hyperbolic 3-manifold is
triangulated by tetrahedrons. However, because of the hyperbolic geometry, the edge between two vertices is not the usual line but rather a geodesics in the hyperbolic geometry. The curvature of this geodesics depends on the hyperbolic structure. In the limit, all geodesics of the tetrahedron meet and one obtains a tree instead of tetrahedrons. Therefore in the limit of large curvature, one obtains a reduction from 3D (=tetrahedrons) to 1D (=tree). Fig. \ref{fig:degen-hyp-tree} visualizes the transition from 2D(triangle) to 1D(tree).

\section{Where do the Quantum Fluctuations Come From?\label{sec:Where-does-fluctuation}}

In a purely geometrical theory, one has to answer this question. It
cannot be shifted to assume the appearance of quantum fluctuations.
Instead we have to understand the root of these quantum fluctuations.
Starting point of our approach is the foliation of the exotic $R^{4}$
by using the Anosov flow. Main point in the argumentation above is
the appearance of the hyperbolic geometry in 3- and 4-dimensional
submanifolds. The foliation can, however, be interpreted differently: a
foliation defines a dynamics at a manifold leading to a splitting
into leafs (the integral curves of the dynamics). Therefore, a tiny
variation in the initial conditions will lead to a strong variation
of the corresponding integral curve. This chaotic behavior is a natural
consequence of the exotic smoothness structure (leading to the non-trivial
$PSL(2,\mathbb{R})-$foliation). For completeness we will describe
this dynamics, called the Anosov flow. For that purpose we consider
the standard basis
\begin{equation}
J=\frac{1}{2}\left(\begin{array}{cc}
1 & 0\\
0 & -1
\end{array}\right)\quad,\, X=\left(\begin{array}{cc}
0 & 1\\
0 & 0
\end{array}\right)\quad,\, Y=\left(\begin{array}{cc}
0 & 0\\
1 & 0
\end{array}\right)\label{generators-SL2R}
\end{equation}
of the Lie algebra $sl(2,\mathbb{R})$ with 
\[
[J,X]=X\quad\quad[J,Y]=-Y\quad\quad[X,Y]=2J
\]
leading to the exponential maps

\[
g_{t}=\exp(tJ)=\left(\begin{matrix}e^{t/2} & 0\\
0 & e^{-t/2}
\end{matrix}\right)\, h_{t}^{*}=\exp(tX)=\left(\begin{matrix}1 & t\\
0 & 1
\end{matrix}\right)\, h_{t}=\exp(tY)=\left(\begin{matrix}1 & 0\\
t & 1
\end{matrix}\right)
\]
defining right-invariant flows on the unit tangent bundle $T_{1}\mathbb{H}=PSL(2,\mathbb{R})$
of the hyperbolic space. The connection to the Anosov flow comes from
the realization that $g_{t}$ is the geodesic flow on $P=T_{1}\mathbb{H}$.
With Lie vector fields being (by definition) left invariant under the action
of a group element, one has that these fields are left invariant under
the specific elements $g_{t}$ of the geodesic flow. This flow goes
over to a surface $M=\mathbb{H}/\Gamma$ defined by a subgroup $\Gamma\subset PSL(2,\mathbb{R})$
with $Q=T_{1}M$. Now the geodesic flow $g_{t}$ acts on the exponential
maps $g_{s},h_{t}^{*},h_{t}$ so that the geodesic flow itself is
invariant, $g_{s}g_{t}=g_{t}g_{s}=g_{s+t}$ , but the other two shrink
and expand: $g_{s}h_{t}^{*}=h_{t\cdot\exp(-s)}^{*}g_{s}$ and $g_{s}h_{t}=h_{t\cdot\exp(s)}g_{s}$.
Then the bundle $TQ$ splits into three subbundles
\[
TQ=E^{+}\oplus E^{0}\oplus E^{-}
\]
where one bundle $E^{+}$ expands, one bundle $E^{-}$ contracts and
one bundle $E^{0}$ is invariant w.r.t. geodesic flow. This property
is crucial for the following discussion. Because of the expanding
behavior of one subbundle, the Anosov flow is the generator of a chaotic
dynamics. Therefore, two geodesics diverge exponentially in this foliation, but this behavior goes over to the holonomies characterizing the geometry.
The transport of a holonomy along two diverging geodesics can lead
to totally different holonomies. Currently this dynamics is deterministic,
i.e. if we choose exactly the same initial condition then we will
end at the state (seen as limit point). This situation changes if
we are unable to choose the initial condition exactly (by choosing
real numbers) but instead we can only choose a rational number where
this rational number is the characterizing property of the state.
Then all initial conditions (represented by all real numbers) in this
class represent the same state but have totally different limit points
of the corresponding dynamics. Now we will describe this dynamics.

\begin{figure}
\includegraphics[scale=0.5]{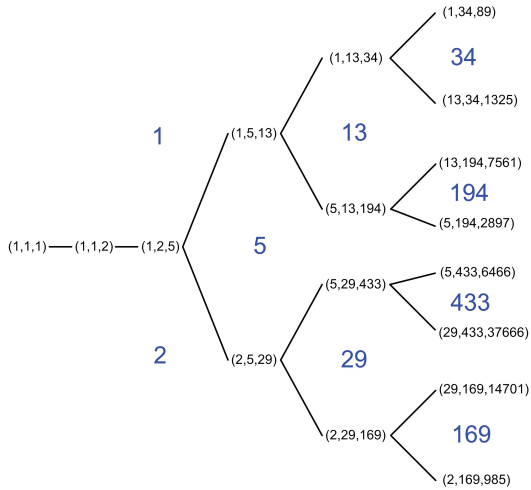}
\caption{\label{fig:binary-tree-of-Markoff}Binary tree of Markoff numbers
as solution of equation (\ref{eq:Markoff-equation}).}
\end{figure}

Starting point is the observable algebra $X(S,SL(2,\mathbb{C})$,
i.e. the space of holonomies $\pi_{1}(S)\to SL(2,\mathbb{C})$ (i.e.
homomorphisms) up to conjugation, see subsection \ref{sub:The-observable-algebra}.
The deformation quantization (see subsection \ref{sub:Drinfeld-Turaev-Quantization})
is the Kauffman bracket skein module. Here we made use of the identity
\[
tr(A)\cdot tr(B)=tr(AB)+tr(AB^{-1})
\]
between two elements $A,B\in SL(2,\mathbb{C})$ (w.r.t. a representation).
Using the group commutator $[A,B]=ABA^{-1}B^{-1}$ one also obtains
\[
2+tr([A,B])=(tr(A))^{2}+(tr(B))^{2}+(tr(AB))^{2}-tr(A)tr(B)tr(AB)
\]
According to deformation procedure, pairs of elements $A,B\in SL(2,\mathbb{C})$
coming from closed curves via the holonomy and fulfilling 
\[
tr([A,B])=\pm2
\]
can be a canonical pair w.r.t. the symplectic structure. The sign
is purely convention and we choose $tr([A,B])=-2$. Then the canonical
pair has to fulfill the equation
\[
(tr(A))^{2}+(tr(B))^{2}+(tr(AB))^{2}-tr(A)tr(B)tr(AB)=0
\]
which can be written in a more familiar form 
\begin{equation}
x^{2}+y^{2}+z^{2}-3xyz=0\label{eq:Markoff-equation}
\end{equation}
by using $3x=tr(A),3y=tr(B),3z=tr(AB)$. Because of the discreteness
of $\pi_{1}(S)$, we have to look for rational solutions of this equation
(Diophantine equation). The solutions of the equation are Markoff
triples forming a binary tree (see Fig. \ref{fig:binary-tree-of-Markoff}).

The set of these elements $A,B$ corresponding to
discrete groups is known to be fractal in nature \cite{Bowditch1998}. 
It is the large class of quasi-Fuchsian
groups having a fractal curve (Julia set) as limit set. Then we
have the desired behavior:\\
\emph{The set of canonical pairs (as measurable variables in the theory)
forms a fractal subset of the space of all holonomies. Then we can
only determine the initial condition up to discrete value (given by
the canonical pair) and the chaotic behavior of the foliation (i.e.
the Anosov flow) makes the limit not predictable.}\\
At the end of this section, one remark about the role of the canonical
pair. It is always possible to construct a classical continuous random
field that has the same probability density as the quantum vacuum
state. Furthermore it is known that a random field can be generated
by a chaotic dynamics. There is, however, a large difference between the
classical random field and a quantum field: there are pairs of not
equally accurate measurable observables (mostly the canonical pairs)
for quantum fields impossible for the classical random fields. With
our approach, we showed the same behavior for the canonical pairs.

\section{Decoherence, Entanglement and Measurement}

Our geometrical approach should also lead to a description of the
measurement process (including the collapse of the wave function).
In section \ref{sec:Wild-embeddings}, we constructed the geometrical
expression for a quantum state given by a wild embedding (the wild
$S^{3}$). The reduction of the quantum state (as linear combination)
to an eigenstate (or the collapse of the wave function) is equivalent
to a reduction of the wild embedding to a tame embedding. Therefore
we need a mechanism to reduce the wild embedding to a tame one. The
construction of the wild $S^{3}$ is strongly related to the Casson
handle. The exoticness of the smooth structure of $R^{4}$ and the
wildness of the $S^{3}$ depend both on the self-intersection of some
disk. If we are able to remove these self-intersections then we will
obtain the desired reduction. According to the discussion in subsection \ref{sub:Small-exotic},
one needs a Casson handle for the cancellation. How many levels
of the Casson handle are needed to cancel the self-intersection? This
question was answered by Freedman \cite{Fre:83}: one needs three
levels (a three-level Casson tower)! At the same time, however, one produces
more self-intersections in the higher levels. Therefore one needs
a little bit more: a Casson tower where a complete Casson handle can be
embedded. Then this Casson handle is able to cancel the self-intersection
and we will obtain a tame embedding or a classical state. As shown
by Freedman \cite{Fre:82} and Gompf/Singh \cite{GompfSingh1984}, one needs a 5-stage
Casson tower so that a Casson handle with the same attaching circle
can be embedded into this 5-stage tower. We obtain a process which
is ''the collapse of the wave function''. What is the cause of
this collapse? As explained above, we cannot choose a single disk to
remove the self-intersections. Instead we have to choose a Casson
tower where each stage is a boundary-connected sum of $S^{1}\times D^{3}$,
i.e. its boundary is the sum $S^{1}\times S^{2}\#S^{1}\times S^{2}\#\cdots$
where the number of components is equal to the number of self-intersections.
So, every piece $S^{1}\times S^{2}$ of the boundary is given by the
identification of the two boundary components for $S^{2}\times[0,1]$.
In section \ref{sec:action-induced-by-top}, we identified this 3-manifold
with the graviton or \emph{the collapse of the wave function is caused
by a gravitational interaction}. The corresponding process is known
as decoherence. In the following we will calculate the minimal decoherence
time for the gravitational interaction. The 5-stage Casson tower can
be also understood as a cobordism between the 3-manifold 
\[
\Sigma_{0}=S^{1}\times S^{2}
\]
 (the $S^{1}$ defines the attaching circle) and a 3-manifold having
the same homology. In case of the simplest Casson tower, it is given
by five complements of the Whitehead link $C(Wh)$ closed by two solid
tori, i.e.
\[
\Sigma_{1}=(D^{2}\times S^{1})\cup C(Wh)\cup C(Wh)\cup C(Wh)\cup C(Wh)\cup C(Wh)\cup(D^{2}\times S^{1})
\]
and this manifold can be very complicated for more complex towers.
Now we will add some geometry to calculate the decoherence time. As
shown by Witten \cite{Wit:89.2,Wit:89.3,Wit:91.2}, the action
\begin{equation}
\intop_{\Sigma_{0,1}}\,^{3}R\sqrt{h}\, d^{3}x=L\cdot CS(\Sigma_{0,1})\label{eq:Witten-relation}
\end{equation}
for every 3-manifold (in particular for $\Sigma_{0}$ and $\Sigma_{1}$
denoted by $\Sigma_{0,1}$) is related to the Chern-Simons action
$CS(\Sigma_{0,1})$. The scaling factor $L$ is related to the volume
by $L=\sqrt[3]{vol(\Sigma_{0,1})}$ and we obtain formally
\begin{equation}
L\cdot CS(\Sigma_{0,1},A)=L^{3}\cdot\frac{CS(\Sigma_{0,1})}{L^{2}}=\intop_{\Sigma_{0,1}}\frac{CS(\Sigma_{0,1})}{L^{2}\cdot vol(\Sigma_{0,1})}\sqrt{h}\, d^{3}x\label{eq:CS-integral-relation}
\end{equation}
by using
\[
L^{3}\cdot vol(\Sigma_{0,1})=\intop_{\Sigma_{0,1}}\sqrt{h}\, d^{3}x
\]
with the (unit) volume $vol(\Sigma_{0,1})$. If $\Sigma_{0,1}$ is
a hyperbolic 3-manifold then the (unit) volume is a topological invariant
which cannot be normalized to 1. Together with 
\[
^{3}R=\frac{3k}{a^{2}}
\]
one can compare the kernels of the integrals of (\ref{eq:Witten-relation})
and (\ref{eq:CS-integral-relation}) to get for a fixed time 
\[
\frac{3k}{a^{2}}=\frac{CS(\Sigma_{0,1})}{L^{2}\cdot vol(\Sigma_{0,1})}\,.
\]
This gives the scaling factor
\begin{equation}
\vartheta=\frac{a^{2}}{L^{2}}=\frac{3\cdot vol(\Sigma_{0,1})}{CS(\Sigma_{0,1})}\label{eq:scaling-CH}
\end{equation}
where we set $k=1$ in the following. The hyperbolic geometry of the
cobordism is best expressed by the metric

\begin{equation}
ds^{2}=dt^{2}-a(t)^{2}h_{ik}dx^{i}dx^{k}\label{eq:FRW-metric}
\end{equation}
also called the Friedmann-Robertson-Walker metric (FRW metric) with
the scaling function $a(t)$ for the (spatial) 3-manifold. Mostow
rigidity enforces us to choose 
\[
\left(\frac{\dot{a}}{a}\right)^{2}=\frac{1}{L^{2}}
\]
in the length scale $L$ of the hyperbolic structure. In the following
we will switch to quadratic expressions because we will determine
the expectation value of the area. A second reason for the consideration
of quadratic expressions is again the hyperbolic structure of $\mathbb{H}^{2}$. We needed this structure for the construction of the foliation
which is given by a polygon in $\mathbb{H}^{2}$. This polygon defines
a compact surface of genus $g>1$. Then the foliation of the polygon
induces a foliation of the small exotic $R^{4}$. The area of the
polygon is mainly the Godbillon-Vey invariant of the foliation. It
is known that foliations of surfaces are given by quadratic differentials
of the form defined below. Here, there are deep connections to trees
and $SL(2,\mathbb{C})$ flat connections, i.e. a tree defines a quadratic
differential and vice versa \cite{Strebel1984,HubbardMasur1979,Wolf1995,DaskalopoulosDostoglouWentworth1998}.

Using the previous equation, we obtain 
\begin{equation}
da^{2}=\frac{a^{2}}{L^{2}}\, dt^{2}=\vartheta\, dt^{2}\label{eq:scale-quadratic-expansion}
\end{equation}
with respect to the scale $\vartheta$. By using the tree of the Casson
handle, we obtain a countable infinite sum of contributions for (\ref{eq:scale-quadratic-expansion}).
Before we start we will clarify the geometry of the Casson handle.
A Casson handle admits a hyperbolic geometry. Therefore the tree corresponding
to the Casson handle must be interpreted as a metric tree with hyperbolic
structure in $\mathbb{H}^{2}$ and metric $ds^{2}=(dx^{2}+dy^{2})/y^{2}$.
The embedding of the Casson handle in the cobordism is given by the
rules
\begin{enumerate}
\item The direction of the increasing levels $n\to n+1$ is identified with
$dy^{2}$ and $dx^{2}$ is the number of edges for a fixed level with
scaling parameter $\vartheta$. 
\item The contribution of every level in the tree is determined by the previous
level best expressed in the scaling parameter $\vartheta$. 
\item An immersed disk at level $n$ needs at least one disk to resolve
the self-intersection point. This disk forms the level $n+1$ but
this disk is connected to the previous disk. So we obtain for $da^{2}|_{n+1}$
at level $n+1$
\[
da^{2}|_{n+1}\sim\vartheta\cdot da^{2}|_{n}
\]
up to a constant. 
\end{enumerate}
By using the metric $ds^{2}=(dx^{2}+dy^{2})/y^{2}$ with the embedding
($y^{2}\to n+1$, $dx^{2}\to\vartheta$) we obtain for the change
$dx^{2}/y^{2}$ along the $x-$direction (i.e. for a fixed $y$) $\frac{\vartheta}{n+1}$.
This change determines the scaling from the level $n$ to $n+1$,
i.e. 
\[
da^{2}|_{n+1}=\frac{\vartheta}{n+1}\cdot da^{2}|_{n}=\frac{\vartheta^{n+1}}{(n+1)!}\cdot da^{2}|_{0}
\]
and after the whole summation (as substitute for an integral in case
of discrete values) we obtain for the relative scaling 
\begin{equation}
a^{2}=\sum_{n=0}^{\infty}\left(da^{2}|_{n}\right)=a_{0}^{2}\cdot\sum_{n=0}^{\infty}\frac{1}{n!}\vartheta^{n}=a_{0}^{2}\cdot\exp\left(\vartheta\right)=a_{0}^{2}\cdot l_{scale}\label{eq:scaling}
\end{equation}
with $da^{2}|_{0}=a_{0}^{2}$. The Chern-Simons invariant for $\Sigma_{0}$
vanishes and we are left with 
\[
CS(\Sigma_{1})=\frac{5}{8}
\]
 and the complements $C(Wh)$ are hyperbolic 3-manifolds with
\[
vol(\Sigma_{1})=5\cdot vol(C(Wh))\approx18.31931...
\]
by using the software Snapea. Finally for the scaling we obtain
\[
\vartheta\approx 87.932688...
\]
and for the time we have to choose
\[
T_{decoherence}=T_{0}\cdot\exp\left(\frac{\vartheta}{2}\right)
\]
using the well-known relation $a_{0}=cT_{0}$ between length and time,
i.e. we see one coordinate along the Casson handle as time axis. The
time $T_{0}$ has to be identified with the Planck time $T_{0}\approx10^{-43}s$
(see section \ref{sec:action-induced-by-top}) so that
\[
T_{decoherence}\approx10^{-24}s
\]
is the minimal decoherence time for the gravitational interaction.

\begin{figure}
\begin{center}
\includegraphics[scale=0.4]{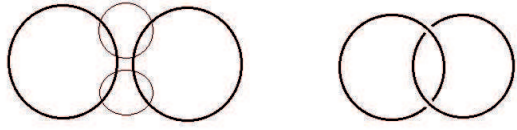}
\caption{two disjoint circles get linked after the application of the skein relation for the area marked by the small circles\label{fig:entangled-circle}}
\end{center}
\end{figure}

Now we also discuss the entanglement which has to be also geometrically
expressed. A quantum state is an element of the skein algebra $K_{t}(S)$
for $S\times[0,1]$. For two disjoint surfaces $S_{0}\sqcup S_{1}$
one has
\[
K_{t}(S_{0}\sqcup S_{1})=K_{t}(S_{0})\otimes K_{t}(S_{1})\;.
\]
Now let us choose a knot $|0\rangle$ in $S_{\text{0}}\times[0,1]$
as element of $K_{t}(S_{0})$ as well a knot $|1\rangle$ in $K_{t}(S_{1})$.
Then $|0\rangle\otimes|1\rangle$ is an element of $K_{t}(S_{0}\sqcup S_{1})$.
Furthermore we can assume that the knots $|1\rangle$ and $|0\rangle$
can be also an element of $K_{t}(S_{0})$ and $K_{t}(S_{1})$, respectively.
Then the element
\[
|0\rangle\otimes|1\rangle+|1\rangle\otimes|0\rangle
\]
exists but now as an element of $K_{t}(S)$ with $S_{0}\sqcup S_{1}\subset S$.
Using the skein relations in $K_{t}(S)$, see Fig. \ref{fig:skein-crossings},
we obtain a linking between the corresponding knots, i.e. $|0\rangle$
and $|1\rangle$ forming a link. Fig. \ref{fig:entangled-circle} visualizes the 
transition from disjoint circles (=disjoint states) to linked circles (=entangled states).
Then entanglement is reduced to a linking!

Next we have to think about the measurement which reduces the entangled
state to one product state. Here we will only present some rough ideas
for the description of the measurement process, but at first we have
to define a measurement device. In this proposal, it is a union of
Casson handles which can be used to unlink two linked components.
At the level of skein algebras, the Casson handle is also given by
elements of a skein algebra (given by closed, knotted curves at the
levels). The particular structure of the Casson handle is not determined
(see also section \ref{sec:Where-does-fluctuation}). Now a given
quantum state is linked to this Casson handle. The limit point of
the Casson handle (i.e. the leafs of the tree) give the result of
the unlinking. All limit points of the Casson handle have a fractal
structure (a Cantor set) expressing our inability to know the outcome
of the measurement. The tree structure of the Casson handle has also
another effect: the limit points are exponentially separated from
each other and can be seen as classical states. With these speculations,
we will close this section.

\section{Some Implications for Cosmology\label{sec:Some-Implications-for.cosmology}}

In the last section we will collect some implications for a cosmological
model. Let us assume the topology $S^{3}\times\mathbb{R}$ for the
spacetime but with an exotic smoothness structure $S^{3}\times_{\theta}\mathbb{R}$.
One can construct this spacetime from the exotic $R^{4}$ by $R^{4}\setminus D^{4}=S^{3}\times_{\theta}\mathbb{R}$.
From previous work, we know:
\begin{itemize}
\item Cosmological anomalies like dark matter and dark energy are (conjecturally) 
rooted in exotic smoothness \cite{TAMBra:2002}.
\item The initial state of the cosmos must be a wild 3-sphere representing
a quantum state \cite{TAMKrol2013}.
\item Then there is an inflationary phase \cite{TAMKrol2014} driven
by a decoherence which can be described by the Starobinsky model.\\
In this model, we have a topological transition from a 3-manifold
$\Sigma_{0}$ to another 3-manifold $\Sigma_{1}$. Both 3-manifolds
are homology 3-spheres. Therefore let us describe this change (a so-called
homology cobordism) between two homology 3-spheres $\Sigma_{0}$ and
$\Sigma_{1}$. The situation can be described by a diagram 
\begin{eqnarray}
\Sigma_{1} & \stackrel{\Psi}{\longrightarrow} & \mathbb{R}\nonumber \\
\phi\downarrow & \circlearrowright & \updownarrow id\label{eq:commuting-diagram}\\
\Sigma_{0} & \stackrel{\psi}{\longrightarrow} & \mathbb{R}\nonumber 
\end{eqnarray}
which commutes. The two functions $\psi$ and $\Psi$ are the Morse
function of $\Sigma_{0}$ and $\Sigma_{1}$, respectively, with $\Psi=\psi\circ\phi$.
The Morse function over $\Sigma_{0,1}$ is a function $\Sigma_{0,1}\to\mathbb{R}$
having only isolated, non-degenerated, critical points (i.e. with
vanishing first derivatives at these points). A homology 3-sphere
has two critical points (located at the two poles). The Morse function
looks like $\pm||x||^{2}$ at these critical points. The transition
$y=\phi(x)$ represented by the (homology) cobordism $M(\Sigma_{0},\Sigma_{1})$
maps the Morse function $\psi(y)=||y||^{2}$ on $\Sigma_{0}$ to the
Morse function $\Psi(x)=||\phi(x)||^{2}$ on $\Sigma_{1}$. The function
$-||\phi||^{2}$ represents also the critical point of the cobordism
$M(\Sigma_{0},\Sigma_{1})$. As we learned above, this cobordism
has a hyperbolic geometry and we have to interpret the function $||\phi(x)||^{2}$
not as an Euclidean form but change it to the hyperbolic geometry so
that
\[
-||\phi||^{2}=-\left(\phi_{1}^{2}+\phi_{2}^{2}+\phi_{3}^{2}\right)\to-
e^{-2\phi_{1}}(1+\phi_{2}^{2}+\phi_{3}^{2})
\]
i.e. we have a preferred direction represented by a single scalar
field $\phi_{1}:\Sigma_{1}\to\mathbb{R}$. Therefore, the transition
$\Sigma_{0}\to\Sigma_{1}$ is represented by a single scalar field
$\phi_{1}:\Sigma_{1}\to\mathbb{R}$ and we identify this field as
the moduli. Finally we interpret this Morse function in the interior
of the cobordism $M(\Sigma_{0},\Sigma_{1})$ as the potential (shifted
away from the point $0$ ) of the scalar field $\phi$ with Lagrangian
\[
L=R+(\partial_{\mu}\phi)^{2}-\frac{\rho}{2}(1-\exp\left(-\lambda\phi\right))^{2}
\]
with two free constants $\rho$ and $\lambda$. For the value $\lambda=\sqrt{2/3}$
and $\rho=3M^{2}$ we obtain the Starobinski model \cite{Starobinski1980}
(by a conformal transformation using $\phi$ and a redefinition of
the scalar field \cite{Whitt1984})
\begin{equation}
L=R+\frac{1}{6M^{2}}R^{2}\label{eq:Starobinski-model}
\end{equation}
with the mass scale $M\ll M_{P}$ much smaller than the Planck mass.
From our discussion above, the appearance of this model is not totally
surprising. It favors a surface to be incompressible (which is compatible
with the properties of hyperbolic manifolds).
\item This inflationary phase is followed by another exponentially increasing
phase leading to a hyperbolic 4-manifold with constant curvature which
is rigid by Mostow rigidity \cite{TAMKrol2014}. Here, we obtained
the global geometry of the spacetime: it is a de Sitter space $SO(4,1)/SO(3,1)$
with a cosmological constant which is the curvature of the spacetime. 
\item This constant curvature can be identified with the cosmological constant
in good agreement with the Planck satellite results \cite{TAMKrol2014}.
The cosmological constant is constant by Mostow rigidity (but now
for the 4-manifold).
\item The topology of the spatial component (seen as 3-manifold) is strongly
restricted \cite{TAMKrol2012} by the smoothness of the spacetime.
\item The inclusion of matter can be done naturally as direct consequence
of exotic smoothness \cite{TAMBrans2015}. 
\item The interior of black holes can be described by exotic smoothness
where the singularity is smoothed out \cite{TAMBrans2012}.
\end{itemize}

\section{Conclusion and Open Questions\label{sec:Conclusion-and-Open-questions}}

Smooth Quantum Gravity, the usage of exotic smoothness structures
on 4-manifolds, are the attempt to obtain a consistent theory of quantum
gravity without any further assumptions. For us, the change of the
smoothness structure is the next step in extending General Relativity,
where non-Euclidean geometry was used to describe gravity and all
accelerations. Then, two different smoothness structures represent
two different physical systems. In particular I think that the standard
smoothness structure represents the case of a spacetime without matter
and non-gravitational fields. In this paper we are going a more radical
way to construct a quantum theory without quantization but by using
purely geometrical ideas from mathematical topics like differential
and geometric topology. The flow of ideas can be simply described
by the following points:
\begin{itemize}
\item An exotic $\mathbb{R}^{4}$ is given by an infinite handlebody (so one needs infinitely many charts) and one finds also the description by an infinite sequence of 3-manifolds together with 4-dimensional cobordisms
connecting them.
\item Every 3-manifold admits a codimension-one foliation which goes over to the 4-dimensional cobordisms. The leaf space of this foliation is an operator algebra with a strong connection to algebraic quantum field theory.
\item The states (as linear functionals in the algebra) depend on knotted
curves and are elements of the Kauffman bracket skein algebra. The
reconstruction of the spatial space gives a wild embedded 3-sphere
which is therefore related to the state, or the quantum state can
be identified with the wild embedding. The classical state is a tame
(i.e. usual) embedding where the deformation quantization of a tame
embedding is a wild embedding.
\item The structure of the operator algebra can be analyzed by the Tomita-Takesaki
modular theory. Then it is possible to construct the quantum action
by using the quantized calculus of Connes.
\item For large scales, one gets the Einstein-Hilbert action. Whereas for
small scales, one obtains a dimensional reduced action.
\item The foliation is given by a hyperbolic dynamics having a chaotic behavior.
For our states, one gets an unpredictable behavior so that the dynamics
can generate the quantum fluctuations.
\end{itemize}
This list shows the current state but there are many open points,
where we list only the most important here:
\begin{itemize}
\item What is the Hamiltonian of the theory? In principle we constructed this operator but have a problem connecting to Loop quantum gravity.
\item What are the states seen as knots? The states are knots but the skein
and Mandelstam identities give a class of knots: the states are conjecturally
the concordance class of knots.
\item Is the state a solution of the Hamiltonian? Here we conjecture that
the concordance class of the knot lies already in the kernel of the
Hamiltonian (therefore it is a solution of the Hamiltonian constraint)
\end{itemize}
A lot is done but there are also many open problems.\\
{\bf Happy Birthday Carl!}
\section*{Acknowledgement}
I have to thank Carl for 20 years of friendship and collaboration as well numerous discussions. Special thanks to Jerzy Kr{\'o}l for our work and many discussions about fundamental problems in math and physics. Now I understand the importance of Model theory.

\section*{Appendix A Casson Handles and Labeled Trees}
\markboth{Appendix A Casson Handles and Labeled Trees}{Appendix A Casson Handles and Labeled Trees}
\addcontentsline{toc}{section}{Appendix A Casson Handles and Labeled Trees}
Let us now consider the basic construction of the Casson handle $CH$.
Let $M$ be a smooth, compact, simply-connected 4-manifold and $f:D^{2}\to M$
a (codimension-2) mapping. By using diffeomorphisms of $D^{2}$ and
$M$, one can deform the mapping $f$ to get an immersion (i.e. injective
differential) generically with only double points (i.e. $\#|f^{-1}(f(x))|=2$)
as singularities \cite{GolGui:73}. But to incorporate the generic
location of the disk, one is rather interesting in the mapping of
a 2-handle $D^{2}\times D^{2}$ induced by $f\times id:D^{2}\times D^{2}\to M$
from $f$. Then every double point (or self-intersection) of $f(D^{2})$
leads to self-plumbings of the 2-handle $D^{2}\times D^{2}$. A self-plumbing
is an identification of $D_{0}^{2}\times D^{2}$ with $D_{1}^{2}\times D^{2}$
where $D_{0}^{2},D_{1}^{2}\subset D^{2}$ are disjoint sub-disks of
the first factor disk. In complex coordinates the plumbing may be
written as $(z,w)\mapsto(w,z)$ or $(z,w)\mapsto(\bar{w},\bar{z})$
creating either a positive or negative (respectively) double point
on the disk $D^{2}\times0$. Consider the pair $(D^{2}\times D^{2},\partial D^{2}\times D^{2})$
and produce finitely many self-plumbings away from the attaching region
$\partial D^{2}\times D^{2}$ to get a kinky handle $(k,\partial^{-}k)$
where $\partial^{-}k$ denotes the attaching region of the kinky handle.
A kinky handle $(k,\partial^{-}k)$ is a one-stage tower $(T_{1},\partial^{-}T_{1})$
and an $(n+1)$-stage tower $(T_{n+1},\partial^{-}T_{n+1})$ is an
$n$-stage tower union of kinky handles $\bigcup_{\ell=1}^{n}(T_{\ell},\partial^{-}T_{\ell})$
where two towers are attached along $\partial^{-}T_{\ell}$. Let $T_{n}^{-}$
be $(\mbox{interior}T_{n})\cup\partial^{-}T_{n}$ and the Casson handle
\[
CH=\bigcup_{\ell=0}T_{\ell}^{-}
\]
is the union of towers (with direct limit topology induced from the
inclusions $T_{n}\hookrightarrow T_{n+1}$). A Casson handle is specified
up to (orientation preserving) diffeomorphism (of pairs) by a labeled
finitely-branching tree with base-point {*}, having all edge paths
infinitely extendable away from {*}. Each edge should be given a label
$+$ or $-$ and each vertex corresponds to a kinky handle; the self-plumbing
number of that kinky handle equals the number of branches leaving
the vertex. The sign on each branch corresponds to the sign of the
associated self plumbing. The whole process generates a tree with
infinite many levels. In principle, every tree with a finite number
of branches per level realizes a corresponding Casson handle. The
simplest non-trivial Casson handle is represented by the tree $Tree_{+}$:
each level has one branching point with positive sign $+$. The reverse
construction of a Casson handle $CH_{\mathcal{T}}$ by using a labeled
tree $\mathcal{T}$ can be found in the appendix A. Let $\mathcal{T}_{1}$
and $\mathcal{T}_{2}$ be two trees with $\mathcal{T}_{1}\subset\mathcal{T}_{2}$
(it is the subtree) then $CH_{\mathcal{T}_{2}}\subset CH_{\mathcal{T}_{1}}$.Given
a labeled based tree $Q$, let us describe a subset $U_{Q}$ of $D^{2}\times D^{2}$.
Now we will construct a $(U_{Q},\partial D^{2}\times D^{2})$ which
is diffeomorphic to the Casson handle associated to $Q$. In $D^{2}\times D^{2}$
embed a ramified Whitehead link with one Whitehead link component
for every edge labeled by $+$ leaving {*} and one mirror image Whitehead
link component for every edge labeled by $-$(minus) leaving {*}.
Corresponding to each first level node of $Q$ we have already found
a (normally framed) solid torus embedded in $D^{2}\times\partial D^{2}$.
In each of these solid tori embed a ramified Whitehead link, ramified
according to the number of $+$ and $-$ labeled branches leaving
that node. We can do that process for every level of $Q$. Let the
disjoint union of the (closed) solid tori in the $n$-th family (one
solid torus for each branch at level $n$ in $Q$) be denoted by $X_{n}$.
$Q$ tells us how to construct an infinite chain of inclusions: 
\[
\ldots\subset X_{n+1}\subset X_{n}\subset X_{n-1}\subset\ldots\subset X_{1}\subset D^{2}\times\partial D^{2}
\]
and we define the Whitehead decomposition $WhC_{Q}=\bigcap_{n=1}^{\infty}X_{n}$
of $Q$. $WhC_{Q}$ is the Whitehead continuum \cite{Whitehead35}
for the simplest unbranched tree. We define $U_{Q}$ to be 
\[
U_{Q}=D^{2}\times D^{2}\setminus(D^{2}\times\partial D^{2}\cup\mbox{closure}(WhC_{Q}))
\]
alternatively one can also write 
\begin{equation}
U_{Q}=D^{2}\times D^{2}\setminus\mbox{cone}(WhC_{Q})\label{eq:UQ-diffeo-CH}
\end{equation}
where $\mbox{cone}()$ is the cone of a space 
\[
cone(A)=A\times[0,1]/(x,0)\sim(x',0)\qquad\forall x,x'\in A
\]
over the point $(0,0)\in D^{2}\times D^{2}$. As Freedman (see \cite{Fre:82}
Theorem 2.2) showed $U_{Q}$ is diffeomorphic to the Casson handle
$CH_{Q}$ given by the tree $Q$.

\section*{Appendix B Thurston Foliation of a 3-Manifold}
\markboth{Appendix B Thurston Foliation of a 3-Manifold}{Appendix B Thurston Foliation of a 3-Manifold}
\addcontentsline{toc}{section}{Appendix B Thurston Foliation of a 3-Manifold}

In \cite{Thu:72} Thurston constructed a foliation of the 3-sphere
$S^{3}$ which depends on a polygon $P$ in the hyperbolic plane $\mathbb{H}^{2}$
so that two foliations are non-cobordant if the corresponding polygons
have different areas. For later usage, we will present the main ideas
of this construction only (see also the book \cite{Tamura1992} chapter
VIII for the details). Starting point is the hyperbolic plane $\mathbb{H}^{2}$
with a convex polygon $K\subset\mathbb{H}^{2}$ having $k$ sides
$s_{1},\ldots,s_{k}$. Assuming the upper half plane model of $\mathbb{H}^{2}$
then the sides are circular arcs. The construction of the foliation
depends mainly on the isometry group $PSL(2,\mathbb{R})$ of $\mathbb{H}^{2}$
realized as rational transformations (and this group can be lifted
to $SL(2,\mathbb{R})$). The followings steps are needed in the construction: 
\begin{enumerate}
\item The polygon $K$ is doubled along one side, say $s_{1}$, to get a
polygon $K'$. The sides are identified by (isometric) transformations
$s_{i}\to s_{i}'$ (as elements of $SL(2,\mathbb{R})$). 
\item Take $\epsilon$-neighborhoods $U_{\epsilon}(p_{i}),U_{\epsilon}(p_{i}')$
with $\epsilon>0$ sufficient small and set 
\begin{align*}
V^{2} & =\left(K\cup K'\right)\setminus\bigcup_{i=1}^{k}\left(U_{\epsilon}(p_{i})\cup U_{\epsilon}(p_{i}')\right)\\
 & =S^{2}\setminus\bigcup_{i=1}^{k}D_{i}^{2}
\end{align*}
having the topology of $V^{2}=S^{2}\setminus\left\{ k\,\mbox{punctures}\right\} $
and we set $P=K\cup K'$. 
\item Now consider the unit tangent bundle $U\mathbb{H}^{2}$, i.e. a $S^{1}-$bundle
over $\mathbb{H}^{2}$ (or the tangent bundle where every vector has
norm one). The restricted bundle over $V^{2}$ is trivial so that
$UV^{2}=V^{2}\times S^{1}$. Let $L,L'$ be circular arcs (geodesics)
in $\mathbb{H}^{2}$ (invariant w.r.t. $SL(2,\mathbb{R})$) starting
at a common point which define parallel tangent vectors w.r.t. the
metrics of the upper half plane model. The foliation of $V^{2}$ is
given by geodesics transverse to the boundary and we obtain a foliation
of $V^{2}\times S^{1}$ (as unit tangent bundle). This foliation is
given by a $SL(2,\mathbb{R})$-invariant smooth 1-form $\omega$ (so
that $\omega=const.$defines the leaves) which is integrable $d\omega\wedge\omega=0$.
($SL(2,\mathbb{R})-$invariant Foliation $\mathcal{F}_{SL}$)
\item With the relation $D^{2}=V^{2}\cup D_{1}^{2}\cup\cdots\cup D_{k-1}^{2}$,
we obtain $D^{2}\times S^{1}=V^{2}\times S^{1}\cup\left(D_{1}^{2}\times S^{1}\right)\cup\cdots\cup\left(D_{k-1}^{2}\times S^{1}\right)$
or the gluing of $k-1$ solid tori to $V^{2}\times S^{1}$ gives a
solid tori. Every glued solid torus will be foliated by a Reeb foliation.
Finally using $S^{3}=(D^{2}\times S^{1})\cup(S^{1}\times D^{2})$
(the Heegard decomposition of the 3-sphere) again with a solid torus
with Reeb foliation, we obtain a foliation on the 3-sphere. 
\end{enumerate}
The construction of this foliation $\mathcal{F}_{Thurston}$ (Thurston
foliation) will be also work for any 3-manifold. Thurston \cite{Thu:72}
obtains for the Godbillon-Vey number 
\[
GV(V^{2}\times S^{1},\mathcal{F}_{SL})=4\pi\cdot vol(P)=8\pi\cdot vol(K)
\]
and 
\begin{equation}
GV(S^{3},\mathcal{F}_{Thurston})=4\pi\cdot Area(P)\label{eq:GV-number-Thurston-foliation-1}
\end{equation}
so that \emph{any real number can be realized by a suitable foliation
of this type}. Furthermore, two cobordant foliations have the same
Godbillon-Vey number (but the reverse is in general wrong). Let $[1]\in H^{3}(S^{3},\mathbb{R})$
be the dual of the fundamental class $[S^{3}]$ defined by the volume
form, then the Godbillon-Vey class can be represented by 
\begin{equation}
\Gamma_{\mathcal{F}_{a}}=4\pi\cdot Area(P)[1]\label{eq:Godbillon-Vey-class-Thurston-foliation}
\end{equation}
The Godbillon-Vey class is an element of the deRham cohomology $H^{3}(S^{3},\mathbb{R})$.
Now we will discuss the general case of a compact 3-manifold carrying
a foliation of the same type like the 3-sphere above. The main idea
of the construction is very simple and uses a general representation
of all compact 3-manifolds by Dehn surgery. Here we will use an alternative
representation of surgery by using the Dehn-Lickorish theorem (\cite{PrasSoss:97}
Corollary 12.4 at page 84). Let $\Sigma$ be a compact 3-manifold
without boundary. There is now a natural number $k\in\mathbb{N}$
so that any orientable 3-manifold can be obtained by cutting out $k$
solid tori from the 3-sphere $S^{3}$ and then pasting them back in,
but along different diffeomorphisms of their boundaries. Moreover,
it can be assumed that all these solid tori in $S^{3}$ are unknotted.
Then any 3-manifold $\Sigma$ can be written as 
\[
\Sigma=\left(S^{3}\setminus\left(\bigsqcup_{i=1}^{k}D_{i}^{2}\times S^{1}\right)\right)\cup_{\phi_{1}}\left(D_{1}^{2}\times S^{1}\right)\cup_{\phi_{2}}\cdots\cup_{\phi_{k}}\left(D_{k}^{2}\times S^{1}\right)
\]
where $\phi_{i}:\partial\left(S^{3}\setminus\left(\bigsqcup_{i=1}^{k}D_{i}^{2}\times S^{1}\right)\right)\to\partial D_{i}^{2}\times S^{1}$
is the gluing map from each boundary component of $\left(S^{3}\setminus\left(\bigsqcup_{i=1}^{k}D_{i}^{2}\times S^{1}\right)\right)$
to the boundary of $\partial D_{i}^{2}\times S^{1}$. This gluing
map is a diffeomorphism of tori $T^{2}\to T^{2}$ (where $T^{2}=S^{1}\times S^{1}$).
The Dehn-Lickorish theorem describes all diffeomorphisms of a surface:
Every diffeomorphism of a surface is the composition of Dehn twists
and coordinate transformations (or small diffeomorphisms). The decomposition
\begin{equation}
S^{3}=\left(V^{2}\times S^{1}\right)\cup\left(D_{1}^{2}\times S^{1}\right)\cup\cdots\cup\left(D_{k-1}^{2}\times S^{1}\right)\cup\left(S^{1}\times D_{k}^{2}\right)\label{eq:decomposition-S3}
\end{equation}
of the 3-sphere can be used to get a decomposition of $\Sigma$ by
\[
\Sigma=\left(V^{2}\times S^{1}\right)\cup_{\phi_{1}}\left(D_{1}^{2}\times S^{1}\right)\cup_{\phi_{2}}\cdots\cup_{\phi_{k}}\left(D_{k}^{2}\times S^{1}\right)
\]
which will guide us to the construction of a foliation on $\Sigma$: 
\begin{itemize}
\item Construct a foliation $\mathcal{F}_{\Sigma,SL}$ on $V^{2}\times S^{1}$
using a polygon $P$ (see above) and 
\item Glue in $k$ Reeb foliations of the solid tori using the diffeomorphisms
$\phi_{i}$. 
\end{itemize}
Finally we get a foliation $\mathcal{F}_{\Sigma,Thurston}$ on $\Sigma$.
According to the rules above, we are able to calculate the Godbillon-Vey
number 
\[
GV(\Sigma,\mathcal{F}_{\Sigma,Thurston})=4\pi\cdot vol(P)
\]
Therefore for any foliation of $S^{3}$, we can construct a foliation
on any compact 3-manifold $\Sigma$ with the same Godbillon-Vey number.
Both foliations $\mathcal{F}_{Thurston}$ and $\mathcal{F}_{\Sigma,Thurston}$
agree for the common submanifold $V^{2}\times S^{1}$ or there is
a foliated cobordism between $V^{2}\times S^{1}\subset\Sigma$ and
$V^{2}\times S^{1}\subset S^{3}$. Of course, $S^{3}$ and $\Sigma$
differ by the gluing of the solid tori but every solid torus carries
a Reeb foliation which does not contribute to the Godbillon-Vey number.

\section*{Appendix C 3-Manifolds and Geometric Structures}
\markboth{Appendix C 3-Manifolds and Geometric Structures}{Appendix C 3-Manifolds and Geometric Structures}
\addcontentsline{toc}{section}{Appendix C 3-Manifolds and Geometric Structures}

\begin{figure}
\centering 
\includegraphics{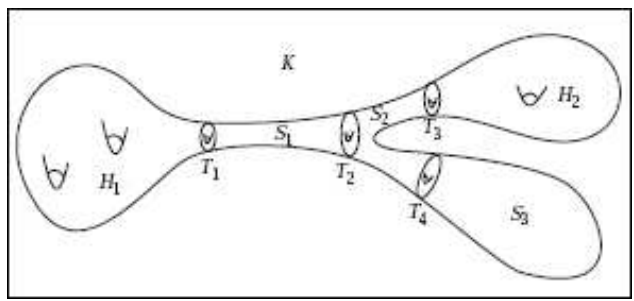}
\caption{Torus (JSJ-) decomposition, $H_{i}$ hyperbolic manifold, $S_{i}$
Graph-manifold, $T_{i}$ Tori. \label{fig:Torus-decomposition}}
\end{figure}

A connected 3-manifold $N$ is prime if it cannot be obtained as a
connected sum of two manifolds $N_{1}\#N_{2}$ neither of which is
the 3-sphere $S^{3}$ (or, equivalently, neither of which is the homeomorphic
to $N$). Examples are the 3-torus $T^{3}$ and $S^{1}\times S^{2}$
but also the Poincare sphere. According to \cite{Mil:62}, any compact,
oriented 3-manifold is the connected sum of a unique (up to homeomorphism)
collection of prime 3-manifolds (prime decomposition). A subset of
prime manifolds are the irreducible 3-manifolds. A connected 3-manifold
is irreducible if every differentiable submanifold $S$ homeomorphic
to a sphere $S^{2}$ bounds a subset $D$ (i.e. $\partial D=S$) which
is homeomorphic to the closed ball $D^{3}$. The only prime but reducible
3-manifold is $S^{1}\times S^{2}$. For the geometric properties (to
meet Thurstons geometrization theorem) we need a finer decomposition
induced by incompressible tori. A properly embedded connected surface
$S\subset N$ is called 2-sided%
\footnote{The `sides' of $S$ then correspond to the components of the complement
of $S$ in a tubular neighborhood $S\times[0,1]\subset N$.%
} if its normal bundle is trivial, and 1-sided if its normal bundle
is nontrivial. A 2-sided connected surface $S$ other than $S^{2}$
or $D^{2}$ is called incompressible if for each disk $D\subset N$
with $D\cap S=\partial D$ there is a disk $D'\subset S$ with $\partial D'=\partial D$.
The boundary of a 3-manifold is an incompressible surface. Most importantly,
the 3-sphere $S^{3}$, $S^{2}\times S^{1}$ and the 3-manifolds $S^{3}/\Gamma$
with $\Gamma\subset SO(4)$ a finite subgroup do not contain incompressible
surfaces. The class of 3-manifolds $S^{3}/\Gamma$ (the spherical
3-manifolds) include cases like the Poincare sphere ($\Gamma=I^{*}$
the binary icosaeder group) or lens spaces ($\Gamma=\mathbb{Z}_{p}$
the cyclic group). Let $K_{i}$ be irreducible 3-manifolds containing
incompressible surfaces then we can $N$ split into pieces (along
embedded $S^{2}$) 
\begin{equation}
N=K_{1}\#\cdots\#K_{n_{1}}\#_{n_{2}}S^{1}\times S^{2}\#_{n_{3}}S^{3}/\Gamma\,,\label{eq:prime-decomposition}
\end{equation}
where $\#_{n}$ denotes the $n$-fold connected sum and $\Gamma\subset SO(4)$
is a finite subgroup. The decomposition of $N$ is unique up to the
order of the factors. The irreducible 3-manifolds $K_{1},\ldots,\, K_{n_{1}}$
are able to contain incompressible tori and one can split $K_{i}$
along the tori into simpler pieces $K=H\cup_{T^{2}}G$ \cite{JacSha:79}
(called the JSJ decomposition). The two classes $G$ and $H$ are
the graph manifold $G$ and the hyperbolic 3-manifold $H$ (see Figure
\ref{fig:Torus-decomposition}). 

The hyperbolic 3-manifold $H$ has a torus boundary $T^{2}=\partial H$,
i.e. $H$ admits a hyperbolic structure in the interior only. In this
paper we need the splitting of the link/knot complement. As shown
in \cite{Budney2006}, the Whitehead double of a knot leads to JSJ
decomposition of the complement into the knot complement and the complement
of the Whitehead link (along one torus boundary of the Whitehead link
complement).

One property of hyperbolic 3-manifolds is central: Mostow rigidity.
As shown by Mostow \cite{Mos:68}, every hyperbolic $n-$manifold
$n>2$ with finite volume has this property: \emph{Every diffeomorphism
(especially every conformal transformation) of a hyperbolic $n-$manifold
with finite volume is induced by an isometry.} Therefore one cannot
scale a hyperbolic 3-manifold and the volume is a topological invariant.
Together with the prime and JSJ decomposition 
\[
N=\left(H_{1}\cup_{T^{2}}G_{1}\right)\#\cdots\#\left(H_{n_{1}}\cup_{T^{2}}G_{n_{1}}\right)\#_{n_{2}}S^{1}\times S^{2}\#_{n_{3}}S^{3}/\Gamma\,,
\]
we can discuss the geometric properties central to Thurstons geometrization
theorem: \emph{Every oriented closed prime 3-manifold can be cut along
tori (JSJ decomposition), so that the interior of each of the resulting
manifolds has a geometric structure with finite volume.} Now, we have
to clarify the term geometric structure's. A model geometry is a simply
connected smooth manifold $X$ together with a transitive action of
a Lie group $G$ on $X$ with compact stabilizers. A geometric structure
on a manifold $N$ is a diffeomorphism from $N$ to $X/\Gamma$ for
some model geometry $X$, where $\Gamma$ is a discrete subgroup of
$G$ acting freely on $X$. t is a surprising fact that there are
also a finite number of three-dimensional model geometries, i.e. 8
geometries with the following models: spherical $(S^{3},O_{4}(\mathbb{R}))$,
Euclidean $(\mathbb{E}^{3},O_{3}(\mathbb{R})\ltimes\mathbb{R}^{3})$,
hyperbolic $(\mathbb{H}^{3},O_{1,3}(\mathbb{R})^{+})$, mixed spherical-Euclidean
$(S^{2}\times\mathbb{R},O_{3}(\mathbb{R})\times\mathbb{R}\times\mathbb{Z}_{2})$,
mixed hyperbolic-Euclidean $(\mathbb{H}^{2}\times\mathbb{R},O_{1,3}(\mathbb{R})^{+}\times\mathbb{R}\times\mathbb{Z}_{2})$
and 3 exceptional cases called $\tilde{SL}_{2}$ (twisted version
of $\mathbb{H}^{2}\times\mathbb{R}$), NIL (geometry of the Heisenberg
group as twisted version of $\mathbb{E}^{3}$), SOL (split extension
of $\mathbb{R}^{2}$ by $\mathbb{R}$, i.e. the Lie algebra of the
group of isometries of 2-dimensional Minkowski space). We refer to
\cite{Scott1983} for the details.

\end{document}